\documentclass{aa}
\usepackage{txfonts}
\usepackage{graphicx}
\usepackage{natbib}
\bibpunct{(}{)}{;}{a}{}{,} 
\usepackage{epsfig}
\usepackage{rotating}

\newcommand{\teff}{\mbox{$T_{\rm eff}$}}
\newcommand{\logg}{\mbox{$\log g$}}
\newcommand{\vsini}{\mbox{$v \sin I$ }}
\newcommand{\mictrb}{\mbox{$\xi_{\rm t}$}}
\newcommand{\mactrb}{\mbox{$v_{\rm mac}$}}

\newcommand{\kms}{\mbox{km\,s$^{-1}$}}

\begin{document}
\title{Spin-orbit angle measurements for six southern transiting planets\thanks{using observations with the high resolution \'echelle spectrograph HARPS mounted on the ESO 3.6\,m (under proposals 072.C-0488, 082.C-0040 \& 283.C-5017), and with the high resolution \'echelle spectrograph CORALIE on the 1.2\,m \textit{Euler} Swiss Telescope, both installed at the ESO La Silla Observatory in Chile. The data is made publicly available at CDS - Strasbourg} 
\\{\large New insights into the dynamical origins of hot Jupiters}}

\author{Amaury H.M.J. Triaud\inst{1}
\and Andrew Collier Cameron\inst{2}
\and Didier Queloz\inst{1}
\and David R. Anderson\inst{3}
\and Micha\"el Gillon\inst{4} 
\and Leslie Hebb\inst{5}
\and Coel Hellier\inst{3}
\and Beno\^it Loeillet\inst{6}
\and Pierre F. L. Maxted\inst{3}
\and Michel Mayor\inst{1}
\and Francesco Pepe\inst{1}
\and Don Pollacco\inst{7}
\and Damien S\'egransan\inst{1}
\and Barry Smalley\inst{3}
\and St\'ephane Udry \inst{1}
\and Richard G. West\inst{8}
\and Peter J. Wheatley\inst{9}
}

\offprints{Amaury.Triaud@unige.ch}

\institute{Observatoire Astronomique de l'Universit\'e de Gen\`eve, Chemin des Maillettes 51, CH-1290 Sauverny, Switzerland
\and School of Physics \& Astronomy, University of St Andrews, North Haugh, St Andrews  KY16 9SS, Fife, Scotland, UK
\and Astrophysics Group, Keele University, Staffordshire ST55BG, UK
\and Institut d'Astrophysique et de G\'eophysique, Universit\'e de Li\`ege, All\'ee du 6 Ao\^ut, 17, Bat. B5C, Li\`ege 1, Belgium
\and Department of Physics and Astronomy, Vanderbilt University, Nashville, TN37235, USA
\and Laboratoire d'Astrophysique de Marseille, BP 8, 13376 Marseille Cedex 12, France
\and Astrophysics Research Centre, School of Mathematics \& Physics, QueenÕs University, University Road, Belfast BT71NN, UK
\and Department of Physics and Astronomy, University of Leicester, Leicester LE17RH, UK
\and Department of Physics, University of Warwick, Coventry CV4 7AL, UK
}

\date{Received date / accepted date}
\authorrunning{Triaud et al.}
\titlerunning{Six Rossiter-McLaughlin effects observed on WASP planets}


\abstract{There are competing scenarii for planetary systems formation and evolution trying to explain how hot Jupiters came to be so close to their parent star. Most planetary parameters evolve with time, making distinction between models hard to do. It is thought the obliquity of an orbit with respect to the stellar rotation is more stable than other parameters such as eccentricity. Most planets, to date, appear aligned with the stellar rotation axis; the few misaligned planets so far detected are massive ( $> 2\, M_\mathrm{J}$).}
{Our goal is to measure the degree of alignment between planetary orbits and stellar spin axes, to detect potential correlation with eccentricity or other planetary parameters and to measure long term radial velocity variability indicating the presence of other bodies in the system.}
{For transiting planets, the Rossiter-McLaughlin effect allows the measurement of the sky-projected angle $\beta$ between the stellar rotation axis and a planet's orbital axis. Using the HARPS spectrograph, we observed the Rossiter-McLaughlin effect for six transiting hot Jupiters found by the WASP consortium. We combine these with long term radial velocity measurements obtained with CORALIE. We used a combined analysis of photometry and radial velocities, fitting models with a Markov Chain Monte Carlo. After obtaining $\beta$ we attempt to statistically determine the distribution of the real spin-orbit angle $\psi$.}
{We found that three of our targets have $\beta$ above $90^{\circ}$: WASP-2b: $\beta = 153^{\circ\,+11}_{\,\,\,\,-15}$, WASP-15b: $\beta = 139.6^{\circ\,+5.2}_{\,\,\,\,-4.3}$ and WASP-17b: $\beta = 148.5^{\circ\,+5.1}_{\,\,\,\,-4.2}$; the other three (WASP-4b, WASP-5b and WASP-18b) have angles compatible with $0^{\circ}$.  There is no dependence between the misaligned angle and planet mass nor with any other planetary parameter. All orbits are close to circular, with only one firm detection of eccentricity on WASP-18b with $e = 0.00848^{+0.00085}_{-0.00095}$. No long term radial acceleration was detected for any of the targets. Combining all previous 20 measurements of $\beta$ and our six and transforming them into a distribution of $\psi$ we find that between about 45 and $85\,\%$ of hot Jupiters have $\psi > 30^{\circ}$.}
{Most hot Jupiters are misaligned, with a large variety of spin-orbit angles. We find observations and predictions using the Kozai mechanism match well. If these observational facts are confirmed in the future, we may then conclude that most hot Jupiters are formed from a dynamical and tidal origin without the necessity to use type I or II migration. At present, standard disc migration cannot explain the observations without invoking at least another additional process.

 }

\maketitle

\section{Introduction}

The formation of close-in gas giant planets, the so-called \textit{hot Jupiters}, has been in debate since the discovery of the first of them, 51\,Peg\,b, by \citet{Mayor:1995p4307}. The repeated observations of these planets in radial velocity and the discovery with HD\,209548b \citep{Charbonneau:2000p5431, Henry:2000p4339} that some of them transit
has produced a large diversity in planetary parameters, such as separation, mass, radius (hence density) and eccentricity. Although more than 440 extrasolar planets have been discovered, of which more than 70 are known to transit, we are still increasing the range of parameters that planets occupy; diversity keeps growing.

\bigskip

While it is generally accepted that close-orbiting gas-giant planets do not form in-situ, their previous and subsequent evolution is still mysterious. Several processes can affect the planet's eccentricity and semi-major axis. Inward migration via angular momentum exchange with a gas disc, first proposed in \citet {Lin:1996p5847} from work by \citet{Goldreich:1980p5872}, is a natural and widely-accepted explanation for the existence of these hot Jupiters. 

Migration alone does not explain the observed distributions in eccentricity and semi-major axis that planets occupy. Alternative mechanisms have therefore been proposed such as the Kozai mechanism  \citep{Kozai:1962p3644, Eggleton:2001p6040, Wu:2003p3071} and planet scattering \citep{Rasio:1996p3680}. These mechanisms can also cause a planet to migrate inwards, and may therefore have a role to play in the formation and evolution of hot Jupiters.
These different models each predict a distribution in semi-major axis and eccentricity. Discriminating between various models is done by matching the distributions they produce to observations. Unfortunately this process does not take into account the evolution with time of the distributions and is made hard by the probable combination of a variety of effects. 

On transiting planets, a parameter can be measured which might prove a better marker of the past history of planets: $\beta$, the projection on the sky of the angle between the star's rotation axis and the planet's orbital axis. 
It is believed that the obliquity (the real spin-orbit angle $\psi$) of an orbit evolves only slowly and is not as much affected by the proximity of the star as the eccentricity  \citep{Hut:1981p2945,Winn:2005p83,Barker:2009p1835}. Disc migration is expected to leave planets orbiting close to the stellar equatorial plane. Kozai cycles and planet scattering should excite the obliquity of the planet and should provide us with a planet population on misaligned orbits with respect to their star's rotation.
\bigskip

As a planet transits a rotating star, it will cause an overall red-shifting of the spectrum if it covers the blue-shifted half of the star and vice-versa on the other side. This is called the Rossiter-McLaughin effect \citep{Rossiter:1924p869, McLaughlin:1924p872}. It was first observed for a planet by \citet{Queloz:2000p247}. Several papers model this effect: \citet{Ohta:2005p631, Gimenez:2006p31,Gaudi:2007p1507}.

Among the 70 or so known transiting planets discovered since 2000 by the huge effort sustained by ground-based transiting planet searches, the Rossiter-McLaughlin (RM) effects have been measured for 20, starting with observations on HD\,209458 by \citet{Queloz:2000p247}. This method has proven itself reliable at giving precise and accurate measurement of the projected spin-orbit angle with its best determination done for HD\,189733b \citep{Triaud:2009p3865}. Basing their analysis on measurements of $\beta$ in 11 systems, 10 of which are coplanar or nearly so, \citet{Fabrycky:2009p1845} concluded that the angle distribution is likely to be bimodal with a coplanar population and an isotropically-misaligned population. 
At that time, the spin-orbit misalignment of XO-3b \citep{Hebrard:2008p226} comprised the only evidence of the isotropic population. Since then, the misalignment of XO-3b has been confirmed by \citet{Winn:2009p3777}, and significant misalignments have been found for HD 80606b \citep{Moutou:2009p2007} and WASP-14b \citep{Johnson:2009p3754}. 
Moreover, retrograde orbital motion has been identified in 
HAT-P-7b \citep{Winn:2009p3712, Narita:2009p5188}. Other systems show indications of misalignment but need confirmation. One such object is WASP-17b \citep{Anderson:2010p5177} which is one of the subjects of the present paper.


\bigskip

The Wide Angle Search for Planets (WASP) project aims at finding transiting gas giants \citep{Pollacco:2006p1500}. Observing the northern and southern hemispheres with sixteen 11\,cm refractive telescopes, the WASP consortium has published more than 20 transiting planets in a large range of period, mass and radius, around stars with apparent magnitudes between 9 and 13. The planet candidates observable from the South are confirmed by a large radial-velocity follow-up using the CORALIE high resolution \'echelle spectrograph, mounted on the 1.2\,m \textit{Euler} Swiss Telescope, at La Silla, Chile. 
As part of our efforts to understand the planets that have been discovered, we have initiated a systematic program to measure the Rossiter-McLaughlin effect in the planets discovered by the WASP survey, in order to measure their projected spin-orbit misalignment angles $\beta$.

\bigskip


In this paper we report the measurement of $\beta$ in six
southern transiting planets from the WASP survey, and analyse
their long term radial velocity behaviour.
In sections 2 and 3 we describe the observations and the methods
employed to extract and analyse the data. In section 4 we report
in detail on the Rossiter-McLaughlin effects observed during transits
of the six systems observed. In sections 5 and 6 we discuss
the correlations and trends that emerge from the study and their
implications for planetary migration models.

\begin{table}
\caption{List of Observations. The date indicates when the first point of the Rossiter-McLaughlin sequence was taken.}\label{tab:obs}
\begin{tabular}{llll}
\hline
\hline
Target	& Date	& Instrument	& Paper\\
\hline
\\
WASP-18b & 2008/08/21	& HARPS & this paper\\
WASP-8b & 2008/10/05	& HARPS & \citet{Queloz:2010p7085}\\
WASP-6b	& 2008/10/07	& HARPS & \citet{Gillon:2009p3869}\\
WASP-4b & 2008/10/08	& HARPS & this paper\\
WASP-5b & 2008/10/10	& HARPS & this paper\\
WASP-2b & 2008/10/15	& HARPS & this paper\\
WASP-15b & 2009/04/27	& HARPS & this paper\\
WASP-17b & 2009/05/22	& CORALIE & this paper\\
WASP-17b & 2009/07/05	& HARPS & this paper\\
\\
\hline

\end{tabular}
\end{table}

\section{The Observations}\label{sec:obs}

\begin{table*}
\caption{Stellar parameters used in our model fitting. The \vsini (stellar spectroscopic rotation broadening) and stellar mass estimates are used as priors in the analysis. $\xi_t$ is the microturbulence. $V_{\rm macro}$ is the macrotrubulence.}\label{tab:SpecParam}
\begin{tabular}{llccccccc}
\hline
\hline
Parameters &\textit{units}&  WASP-2 (a,b)           & WASP-4 (c)            & WASP-5 (c)             & WASP-15 (d)           & WASP-17 (a,e)          & WASP-18 (f)\\
\hline
           \\
\multicolumn{2}{l}{Spectral Type} &K1&G8&G5&F7&F4&F6\\
\\
\teff  \, &K   & 5150 $\pm$ 80    & 5500 $\pm$ 100   & 5700 $\pm$ 100    & 6300 $\pm$ 100   & 6650 $\pm$ 80    & 6400 $\pm$ 100 \\
$B-V$&mag&	0.86 $\pm$ 0.05 & 0.73 $\pm$ 0.05 & 0.66 $\pm$ 0.05 &0.48  $\pm$ 0.05& 0.38  $\pm$ 0.05& 0.45 $\pm$ 0.05\\
\logg     &dex & 4.40 $\pm$ 0.15    & 4.5 $\pm$ 0.2      & 4.5 $\pm$ 0.2       & 4.35 $\pm$ 0.15    & 4.45 $\pm$ 0.15    & 4.4 $\pm$ 0.15    \\
{[Fe/H]}   &dex&$-$0.08 $\pm$ 0.08  & $-0.03$ $\pm$ 0.09 & +0.09 $\pm$ 0.09    &$-$0.17 $\pm$ 0.11  &$-$0.19 $\pm$ 0.09  &   0.00 $\pm$ 0.09 \\
\\
log\,$R^\prime_\mathrm{HK}$&dex &$-$4.84 $\pm$ 0.10& $-$4.50 $\pm$ 0.06 &$-$4.72 $\pm$ 0.07&$-$4.86 $\pm$ 0.05&-& $-$4.85 $\pm$ 0.02 \\
\\
\mictrb \,  &km\,s$^{-1}$ & 0.9 $\pm$ 0.1  & 1.1 $\pm$ 0.2 & 1.2 $\pm$ 0.2  & 1.4 $\pm$ 0.1 & 1.7 $\pm$ 0.1 & 1.6 $\pm$ 0.1 \\
$V_{\rm macro}$  &km\,s$^{-1}$        & 1.6 $\pm$ 0.3  & 2.0 $\pm$ 0.3& 2.0 $\pm$ 0.3& 4.8 $\pm$ 0.3& 6.2 $\pm$ 0.3 & 4.8 $\pm$ 0.3 \\
\vsini  &km\,s$^{-1}$   & 1.6 $\pm$ 0.7 & 2.0 $\pm$ 1.0 & 3.5 $\pm$ 1.0  & 4.0 $\pm$ 2.0     & 9.8 $\pm$ 0.5 & 11.0 $\pm$ 1.5 \\
\\
$M_{\star}$ &$\mathrm{M}_{\odot}$ & $0.84 \pm 0.11$ &$ 0.93 \pm 0.05$ & $1.00\pm 0.06$& $1.18 \pm 0.12 $ &$1.2 \pm 0.12 $& $1.24 \pm 0.04$\\
\\
\hline
\end{tabular}
\note{references:  (a) this paper, (b) \citet{Cameron:2007p2879}, (c) \citet{Gillon:2009p1630}, (d) \citet{West:2009p2783}, (e) \citet{Anderson:2010p5177}, (f) \citet{Hellier:2009p3885}}
\end{table*}

In order to determine precisely and accurately the angle $\beta$, we need to obtain radial velocities during planetary transits at a high cadence and high precision. We therefore observed with the high resolution \'echelle spectrograph HARPS, mounted at the La Silla 3.6\,m ESO telescope. The magnitude range within which planets are found by the SuperWASP instruments allows us to observe each object in adequate conditions. 
For the main survey proposal 082.C-0040, we selected as targets the entire population of transiting planets known at the time of proposal submission to be observable from La Silla during Period 82, i.e. WASP-2b, 4b, 5b, 6b, 8b and 15b. The results for WASP-6b are presented separately by \citet{Gillon:2009p3869} and for WASP-8b by  \citet{Queloz:2010p7085}. Two targets were added in separate proposals. A transit of WASP-18b was observed during GTO time (072C-0488) of the HARPS consortium allocated to this planet because of its short and eccentric orbit. During the long-term spectroscopic follow-up of WASP-17b undertaken for the discovery paper \citep{Anderson:2010p5177}, three CORALIE measurements fell during transit showing a probably retrograde orbit. Observations of the Rossiter-McLaughlin with CORALIE confirmed the conclusions of \citet{Anderson:2010p5177}, and a follow-up DDT proposal (283.C-5017) was awarded time on HARPS. 



The strategy of observations was to take two high precision HARPS points the night before transit and the night after transit. 
The radial-velocity curve was sampled densely throughout the transit, beginning
90 minutes before ingress and ending 90 minutes after egress. The data taken before
ingress and after egress allow any activity-related offset in the effective
velocity of the system's centre of mass to be determined for the night of observation.
In addition, radial velocity data from the high resolution \'echelle spectrograph CORALIE mounted on the Swiss 1.2\,m \textit{Euler} Telescope, also at La Silla was acquired to help search for a long term variability in the the periodic radial velocity signal.




All our HARPS observations have been conducted in the OBJO mode, without simultaneous Thorium-Argon spectrum. The velocities are estimated by a Thorium-Argon calibration at the start of the night. HARPS is stable within 1\,m\,s$^{-1}$ across a night. This is lower than our individual error bars and leads to no contamination of the Th-Ar lamp onto the stellar spectrum easing spectral analysis.

\section{The Data Analysis}\label{sec:analysis}

\subsection{Radial-velocity extraction}

The spectroscopic data were reduced using the online Data Reduction Software (DRS) which comes with HARPS. The radial velocity information was obtained by removing the instrumental blaze function and cross-correlating each spectrum with one of two masks. This correlation is compared with the Th-Ar spectrum acting as a reference; see \citet{Baranne:1996p1069}, \citet{Pepe:2002p1068} \& \citet{Mayor:2003p4493} for details. Recently the DRS was shown to achieve remarkable precision \citep{Mayor:2009p4452} thanks to a revision of the reference lines for Thorium and Argon by \citet{Lovis:2007p1122}. 
Stars with spectral type earlier than G9 were reduced using the G2 mask, while those of K0 or later were cross-correlated with the K5 mask. A similar software package is used for CORALIE data. A resolving power $R=110\,000$ for HARPS yields a cross-correlation function (CCF) binned in $0.25$\,km\,s$^{-1}$ increments, while for CORALIE, with a lower resolution of 50\,000, we used $0.5$\,km\,s$^{-1}$. 
The CCF window was adapted to be three times the size of the full width at half maximum (FWHM) of the CCF. 

All our past and current CORALIE data on the stars presented here were reprocessed after removal of the instrumental blaze response, thereby changing slightly some radial velocity values compared to those already published in the literature. Correcting this blaze is important for extracting the correct RVs for the RM effect. The uncorrected blaze created a slight systematic asymmetry in the CCF that was translated into a bias in radial velocities. 

$1\,\sigma$ error bars on individual data points were estimated from photon noise alone. HARPS is stable long term within 1\,m\,s$^{-1}$ and CORALIE at less than 5\,m\,s$^{-1}$. These are smaller than our individual error bars and thus have not been taken into account. 

\subsection{Spectral analysis}\label{sec:SpecAn}

Spectral analysis is needed to determine the stellar atmospheric parameters from which limb darkening coefficients can be inferred. We carried out new analyses for two of the target stars, WASP-2 and WASP-17, whose previously-published spectroscopic parameters were of low precision. For our other targets, the atmospheric parameters were taken from the literature, notably the stellar spectroscopic rotation broadening \vsini \footnote{throughout this paper we use the symbol $I$ to denote the inclination of the stellar rotation axis to the line of sight, while $i$ represents the inclination of the planet's orbital angular momentum vector to the line of sight}.

The individual HARPS spectra can be co-added to form an overall spectrum above S/N $\sim1:100$, suitable for photospheric analysis which was performed using the {\sc uclsyn} spectral synthesis package  \citep{Smith92, Smalley01} and {\sc atlas9} models without convective overshooting \citep{Castelli:1997p4626} and the same method as described in many discovery papers published by the WASP consortium (eg: \citet{Wilson:2008p2374}).

The stellar rotational \vsini is determined by fitting the profiles of
several unblended Fe~{\sc i} lines. The instrumental FWHM was determined to be $0.065 \AA$  from the telluric lines around $6300 \AA$. 

For WASP-2, a value for macroturbulence (\mactrb) of
1.6 $\pm$ 0.3~\kms\  was adopted \citep{Gray:2008p4677}. A best fitting value of $\vsini = 1.6
\pm 0.7$\,km\,s$^{-1}$ was obtained.
On WASP-17, a value for macroturbulence (\mactrb) of
6.2 $\pm$ 0.3~\kms\ was used  \citep{Gray:2008p4677}. The analysis gives a  best fitting value of \vsini$= 9.8 
\pm 0.5$\,km\,s$^{-1}$. The error on $\mactrb$  is taken from the scatter around fit to \citet{Gray:2008p4677} and is propagated to the \vsini.




$B-V$ were estimated from the effective temperature and used in the calculations of the $\log R^\prime_\mathrm{HK}$ \citep{Noyes:1984p6853,Santos:2000p6686,Boisse:2009p1077}.  Errors refer to the photon noise: they do not include systematic effects likely to arise and affect low values of $\log R^\prime_\mathrm{HK}$ due to the low signal to noise in the blue orders.
WASP-17 does not have a value since this stellar activity indicator is only calibrated for $B-V \in [0.44,1.20]$.

All stellar parameters, used as well as derived, are presented in Table~\ref{tab:SpecParam}.

\subsection{Model fitting}

The extracted radial velocity data was fitted simultaneously with the transit photometry available at the time of analysis. Three models are adjusted to the data: a Keplerian radial velocity orbit \citep{Hilditch:2001p773}, a photometric planetary transit \citep{Mandel:2002p768}, and a spectroscopic transit, also known as Rossiter-McLaughlin effect \citep{Gimenez:2006p31}. This combined approach is very useful for taking into account all of the possible contributions to the uncertainties due to correlations among all relevant parameters. 
A single set of parameters describes both the photometry and the  radial velocities. We use a Markov Chain Monte Carlo (MCMC) approach to optimize the models and estimate the uncertainties of the fitted parameters. The fit of the model to the data is quantified using the $\chi^2$ statistic.

The code is described in detail by  \citet{Triaud:2009p3865}, has been used several times (eg: \citet{Gillon:2009p3869}) and is similar to the code described in \citet{Cameron:2007p2879}. 

We fitted up to 10 parameters, namely the depth of the primary transit $D$, the radial velocity (RV) semi-amplitude $K$, the impact parameter $b$, the transit width $W$, the period $P$, the epoch of mid-transit $T_0$, $e \cos \omega$, $e \sin \omega$, $V\sin I\,\cos \beta$, and $V\sin I\,\sin \beta$. Here $e$ is the eccentricity and $\omega$ the angle between the line of sight and the periastron, $V\sin I$ is the sky-projected rotation velocity of the star\footnote{we make a distinction between \vsini and $V\sin I$: \vsini is the value extracted from the spectral analysis, the stellar spectroscopic rotation broadening, while  $V\sin I$ denotes the result of a Rossiter-McLaughlin effect fit. Both can at times be different. Each, although caused by the same effect, is independently measured making the distinction worthwhile.} while $\beta$ is the sky-projected angle between the stellar rotation axis \citep{Hosokawa:1953p2002, Gimenez:2006p31} and the planet's orbital axis\footnote{$\beta = -\lambda$, another notation used in the literature for the same angle.}. 

These parameters have been chosen to reduce correlations between then. The use of uncorrelated parameters allows to explore parameter space more efficiently since the correlation length between jumps is smaller.
Eccentricity and periastron angle were paired as were $V\sin I$ and $\beta$. This breaks a correlation between them 
(the reader is invited to compare Figs.~\ref{fig:WASP4dis}d \& \ref{fig:betas} for a clear illustration for choosing certain jump parameters as opposed to others). This way we also explore solutions around zero more easily: $e\cos \omega$ and $e \sin \omega$ move in the ]-1,1[ range while $e$ could only be floating in ]0,1[. For exploring particular solutions such as a circular orbit, parameters can be fixed to certain values.

We caution that, as noted by \citet{Ford:2006p7023} that the choice of $e\cos\omega$ and $e\sin\omega$ as jump variables implicitly imposes a prior that is proportional to $e$. This approach thus has a tendency to yield a higher eccentricity than would be obtained with a uniform prior, in cases when $e$ is poorly-constrained by the data. A similar argument applies to the use of $V\sin I \cos \beta$ and $V \sin I \sin \beta$ as jump variables, in cases where the impact parameter is low and there is a strong degeneracy between $V\sin I\,$and $\beta$ in modelling the Rossiter-McLaughlin effect. In such cases, however, the tendency to overestimate $V \sin I$ from the Rossiter-McLaughlin effect can effectively be curbed by imposing an independent, spectroscopically-determined $v \sin I$ prior on $V \sin I$, as we have done here.

In addition to the physical free floating parameters, we need to use one $\gamma$ velocity for each RV set and one normalisation factor for each lightcurve as adjustment parameters. These are found by using optimal averaging and optimal scaling. $\gamma$ velocities represent the mean radial velocity of the star in space with respect to the barycentre of the Solar System. 
Since our analysis had many datasets, the results for these adjustment parameters have been omitted, not adding anything to the discussion.

During these initial analyses we also fitted an additional acceleration in the form of an RV drift $\dot{\gamma}$ but on no occasion was it significantly different from zero. We therefore assumed there was no drift for any of our objects. We will give upper limits for each star in the following sections.


The MCMC algorithm perturbs the fitting parameters at each step $i$ with a simple formula:
\begin{equation}\label{eq:a}
P_{i ,j}= P_{i-1,j} + f\,\sigma_{P_j}\,G(0,1)
\end{equation}
where $P_j$ is a free parameter, $G$ is a Gaussian random number of unit standard deviation and zero mean (meaning a Gaussian prior on each parameter), while $\sigma$ is the step size for each parameter. A factor $f$ is used to control the chain and ensures that $25\,\%$ of steps are being accepted via a Metropolis-Hastings algorithm, as recommended in \citet{Tegmark:2004p793} to give an optimal exploration of parameter space. 

The step size is adapted by doing several initial analyses. They are adjusted to produce as small a correlation length as possible. Once the value is chosen, it remains fixed. Only $f$ fluctuates. 

A \textit{burn-in} phase of 50\,000 accepted steps is used to make the chain converge. This is detected when the correlation length of each parameter is small and that the average $\chi^2$ does not improve anymore \citep{Tegmark:2004p793}. Then starts the real chain, of 500\,000 accepted steps, from which results will be extracted. This number of steps is used as a compromise between computation time and exploration. 
Statistical tests, notably by comparing $\chi^2$ are used to estimate the significance of the results.

Bayesian penalties acting as prior probability distribution can be added to $\chi^2$ to account for any prior information that we might have on any fitted or derived parameter. Stellar mass $M_{\star}$ can notably be inserted via a prior in the MCMC in order to propagate its error bars on the planet's mass. We also inserted the \vsini found by spectral analysis as priors in some of our fits to control how much the fit was dependent on them and the resulting value of $V\sin I$ and whether this influenced the fitted value of $\beta$. Because of the random nature of an MCMC, sometimes a step with an impact parameter close to zero is taken. This can cause $V\,\sin\,I$ to wander to unphysical values because of the degeneracy between $V\,\sin\,I$ and $\beta$ at low impact parameters. This is controlled by imposing a prior. The prior values are in Table \ref{tab:SpecParam}.

We use a quadratic limb-darkening law with fixed values for the two limb darkening coefficients appropriate to the stellar effective temperature. They were extracted for the photometry from tables published in \citet{Claret:2000p364}. For the radial velocity (the Rossiter-McLaughlin effect is also dependent on limb darkening) we use values for the $V$ band. \citet{Triaud:2009p3865} showed that HARPS is centred on the $V$ band. The coefficients were chosen for atmospheric parameters close to those presented in Table~\ref{tab:SpecParam}.

\begin{figure*}
\centering                     
\includegraphics{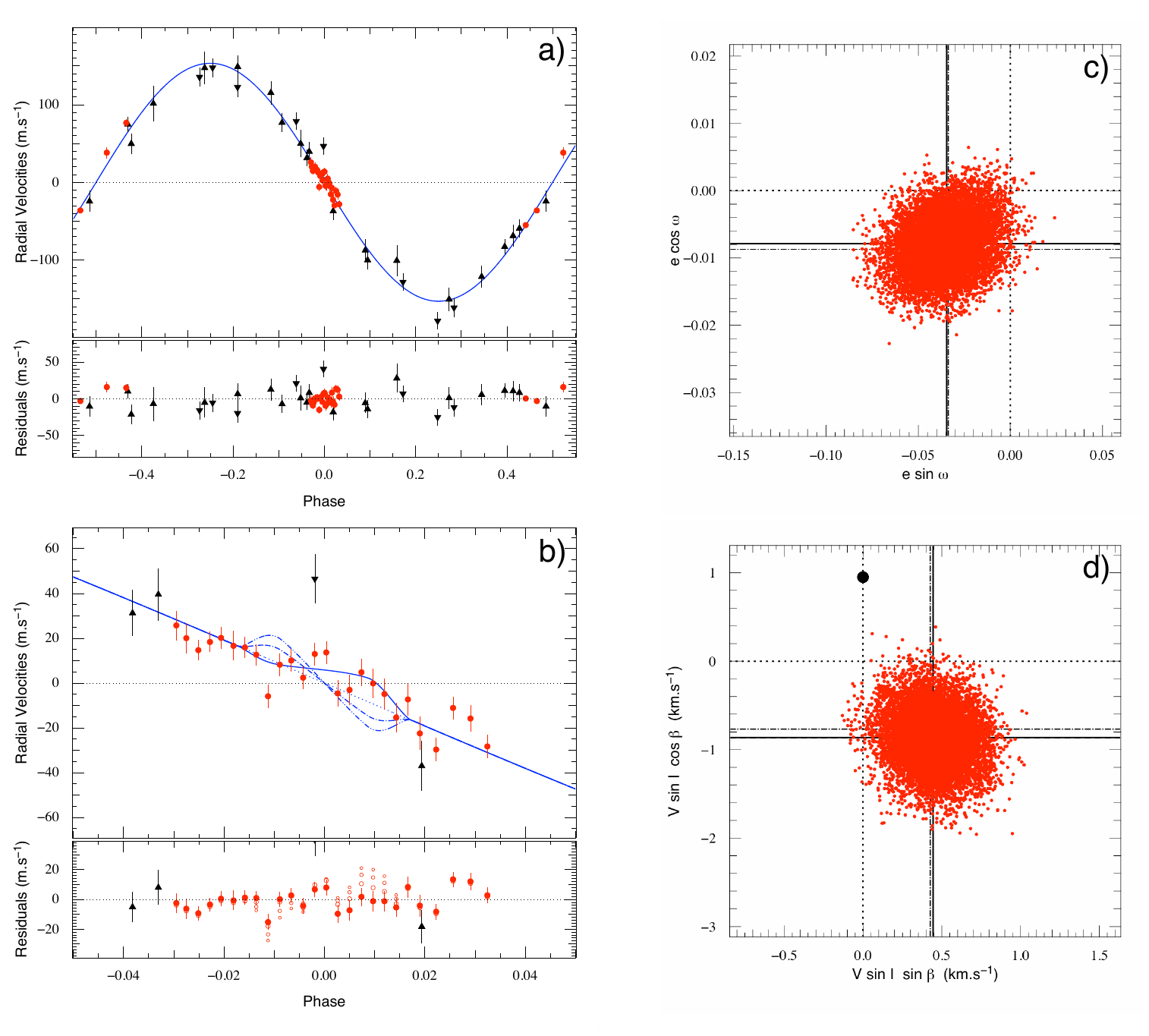}
\caption{
 Fit results for WASP-2b. 
\textit{a)} Overall Doppler shift reflex motion of the star due to the planet and residuals.
\textit{b)} Zoom on the Rossiter-McLaughlin effect and residuals. 
Black inverted triangles are SOPHIE data, black triangles represent CORALIE points, red dots show the HARPS data. The best fit model is also pictured as a plain blue line. 
In addition to our best model found with $V\sin I = 0.99$\,km\,s$^{-1}$  we also present models with no RM effect plotted as a dotted blue line, RM effect with $\beta=0$  and $V\sin I = 0.9$\,km\,s$^{-1}$ drawn with a dashed-dotted blue line and RM effect with $\beta=0$ and $V\sin I = \vsini = 1.6$\,km\,s$^{-1}$ pictured with a dashed-double dotted blue line. In the residuals, the open symbols represent in the values with the size of the circle decreasing with the likelyhood of the model.
\textit{c)} Posterior probability distribution issued from the MCMC showing the distribution of points between $e \cos \omega$ and $e \sin \omega$.
\textit{d)} Posterior probability distribution issued from the MCMC showing the distribution of points between $V\sin I\,\cos \beta$ and $V\sin I\,\sin \beta$.  The \textbf{black disc} shows where the distribution would be centred only changing to $\beta = 0$.
The dotted line shows where zero is.
The straight lines represent the median of the distribution, the dashed lines plot the position of the average values, the dash-dotted lines indicate the values with the lowest $\chi^2$ (some lines can overlap). The size of boxes \textit{c)} and \textit{d)} represents 7 times the $1\,\sigma$ distance on either side of the median. 
}\label{fig:WASP2dis}
\end{figure*}

\subsection{Extracting the results}\label{subsec:results}

For each star, we performed four analyses, each using a MCMC chain with 500\,000 accepted steps:
\begin{itemize}
\item 1. a prior is imposed on $V\sin I$, eccentricity is fixed to zero; 
\item 2. no prior on $V\sin I$, eccentricity is fixed to zero; 
\item 3. a prior is imposed on $V\sin I$, eccentricity is let free;
\item 4. no prior on $V\sin I$, eccentricity is let free. 
\end{itemize}

This is to assess the sensitivity of the model parameters to a small but uncertain orbital eccentricity and to the \vsini value found by spectral analysis which, as demonstrated in \citet{Triaud:2009p3865}, can seriously affect the fitting of the Rossiter-McLaughlin effect. The comparative tables holding the results of these various fits are available in the appendices to support the conclusions we reach while allowing readers to form their own opinion. In addition, we also conducted control chains fixing the parameters controlled by the photometry in order to check whether this was a limiting source of errors in the determination of our most important parameter: $\beta$. The results from these chains are in the appendices as well. Although different in their starting hypotheses all the chains are also useful at checking their respective convergence. Our final results are presented in Table~\ref{tab:params}. 

The best solution is found in the best of the four fits by comparing $\chi^2$ and using Ockham's principle of minimising the number of parameters for similar results: for fits with similar $\chi^2_\mathrm{reduced}$ we usually choose a circular solution with no prior on $V\sin I$. Results are extracted from the best fit by taking the median of the posterior probability distribution for each parameter, determined from the Markov chain. Errors bars are estimated from looking at the extremes of the distribution comprising the 68.3\,\% confidence region of the accepted steps. The best solution is not taken from the lowest $\chi^2$ as it is dependent on the sampling and chance encounter of a - small - local minimum. Scatter plots will be presented with the positions of the best $\chi^2$, the average and the median for illustration.

In the following section and in tables, several statistical values are used: $\chi^2$ is the value found for all the data, while $\chi^2_\mathrm{RV}$ gives the value of $\chi^2$ solely for the radial velocities. The reduced $\chi^2$ for the radial velocities, denoted by $\chi^2_\mathrm{reduced}$,  is used to estimate how well a model fits the data and to compare various fits and their respective significance. In addition we will also use the residuals, denoted as $O-C$. These estimates are only for radial velocities. The results from photometry are not mentioned since they are not new. They are only here to constrain the shape of the Rossiter-McLaughlin effect.

When giving bounds, for eccentricity and long term radial velocity drift, we quote the 95\% confidence interval for exclusion.

\section{The Survey Results}\label{sec:survey}








\begin{figure*}
\centering                     
\includegraphics{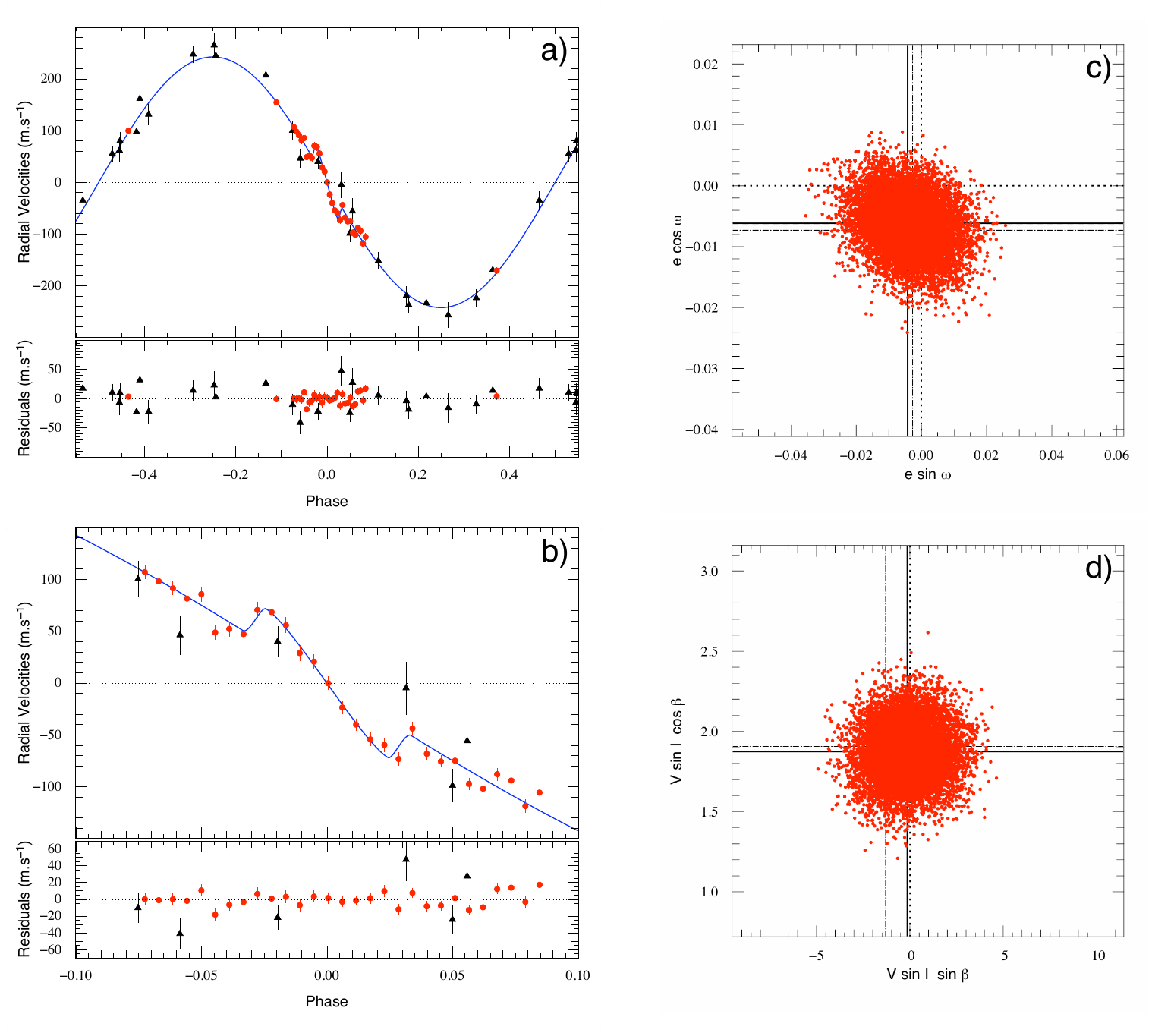}
\caption{
Fit results for WASP-4b. 
\textit{Nota Bene}: Legend similar to the legend in Fig.\ref{fig:WASP2dis}. 
}\label{fig:WASP4dis}


\end{figure*}

\subsection{WASP-2b}\label{subsec:WASP2}




A sequence of 26 RV measurements was taken on WASP-2 using HARPS on 2008 October 15, with additional observations made outside transit as given in the journal of observations presented in the appendices. The cadence during transit was close to a point every 430s. 
The average photon noise error of that sequence is $5.7$\,m\,s$^{-1}$. 
We made additional observations with CORALIE to refine the orbital
solution obtained by \citet{Cameron:2007p2879} using the SOPHIE instrument
on the 1.93\,m telescope at Observatoire de Haute-Provence, and to look for
long-term variability of the orbit.
20 measurements were taken with a mean precision of  $13.9$\,m\,s$^{-1}$ over close to 11 months between 2008 October 25 and 2009 September 23. All the RV data is available in the appendices along with exposure times.

To establish the photometric ephemeris and the transit geometry, we fitted the photometric datasets of  \citet{Cameron:2007p2879} (3 seasons by SuperWASP in the unfiltered WASP bandpass),  \citet{Charbonneau:2007p1458} (a $z$ band $Keplercam$ lightcurve) and  \citet{Hrudkova:2009p1459} (a William Herschel Telescope \textit{AG2} $R$ band transit curve).
\bigskip

WASP-2b's data were fitted with up to 10 free parameters plus 8 independent adjustment parameters: three $\gamma$ velocities for the three RV data sets 
and  five normalisation factors for photometry. 
 This sums up to 58 RV measurements and 8951 photometric observations. 



$\chi^2_\mathrm{reduced}$ does not improve significantly between circular and eccentric models. We therefore impose a circular solution. The presence of a prior on $V\sin I$ does not affect the results. We find $V\sin I = 0.99 ^{+0.27}_{-0.32}$
\,km\,s$^{-1}$ in accordance with the \vsini value found in section \ref{sec:SpecAn}. The fit delivers $\beta = 153^{\circ\,+11}_{\,\,\,\,-15}$. The overall root-mean-square (RMS) scatter of the spectroscopic residuals about the fitted model is 11.73\,m\,s$^{-1}$. During the HARPS transit sequence these residuals are at 6.71\,m\,s$^{-1}$.

Fig.~\ref{fig:betas} shows the resulting distribution as $V\sin I$ vs. $\beta$. We detect $V\,\sin\,I$ significantly above zero with confidence interval showing that 99.73\% ($3\,\sigma$) of the posterior probability function has $V\sin I > 0.2$\,km\,s$^{-1}$ while $\beta > 77.26^{\circ}$.
We have computed  6 additional chains in order to test the strength of our conclusions. Table~\ref{tab:WASP1comp} shows the comparison between the various fits; we invite the reader to refer to it as only important results are given in the text. 


In all cases, eccentricity is not detected being below a $3\,\sigma$ significance from circular which is likely affected from the poor coverage of the phase by the HARPS points. 
Circular solutions are therefore adopted. We fix the eccentricity's upper limit to $e < 0.070$. In addition no significant long term drift was detected in the spectroscopy: $| \dot{\gamma} | < 36 $\,m\,s$^{-1}$\,yr$^{-1}$.


Using the spectroscopically-determined $\vsini$ value of 1.6\,km\,s$^{-1}$ and forcing $\beta$ to zero, $\chi^2_\mathrm{reduced}$  changes from $2.14\pm0.27$ to $3.49\pm0.39$, clearly degrading the solution. We are in fact $7.6\,\sigma$ away from the best-fitting solution, therefore excluding an aligned system with this large a $V\sin I$. This is also excluded by comparison to a fit with a flat RM effect at the $6.7\,\sigma$. Similarly, a fit with an imposed $V\sin I = 0.9$\,km\,s$^{-1}$ and aligned orbit is found $5.6\,\sigma$ from our solution.
On Fig~\ref{fig:WASP2dis}b, we have plotted the various models tested and their residuals so as to give a visual demonstration of the degradation for each of the alternative solutions.


\subsection{WASP-4b}\label{subsec:WASP4}


We obtained a RM sequence of WASP-4b with HARPS on 2008 October 8; other, out of transit, measurements are reported in the journal of observations given in the appendices. The RM sequence comprises 30 data points, 13 of which are in transit, taken at a cadence of 630\,s$^{-1}$ with a mean precision of $6.4$\,m\,s$^{-1}$. 
The spectrograph CORALIE continued monitoring WASP-4 and we add ten radial velocity measurements to the ones published in \citet{Wilson:2008p2374}. These new data were observed around the time of the HARPS observations, about a year after spectroscopic follow-up started. 

In photometry we gathered 2 timeseries in the WASP bandpass from \citet{Wilson:2008p2374} and an $R$ band \textit{C2 Euler} transit plus a \textit{VLT/FORS2} $z$ band lightcurve obtained from \citet{Gillon:2009p1630} to establish the transit shape and timing.

The WASP-4b data were fitted with up to 10 free parameters to which 6 adjustment parameters were added: two $\gamma$ velocities for RVs and four normalisation factors for the photometry. In total, this represents 56 radial velocity points and 9989 photometric measurements. \citet{Gillon:2009p1630} let combinations of limb darkening coefficients free to fit the high precision \textit{VLT} curve. We used and fixed our coefficients on their values.  

\begin{figure*}
\centering                     
\includegraphics[width=16cm]{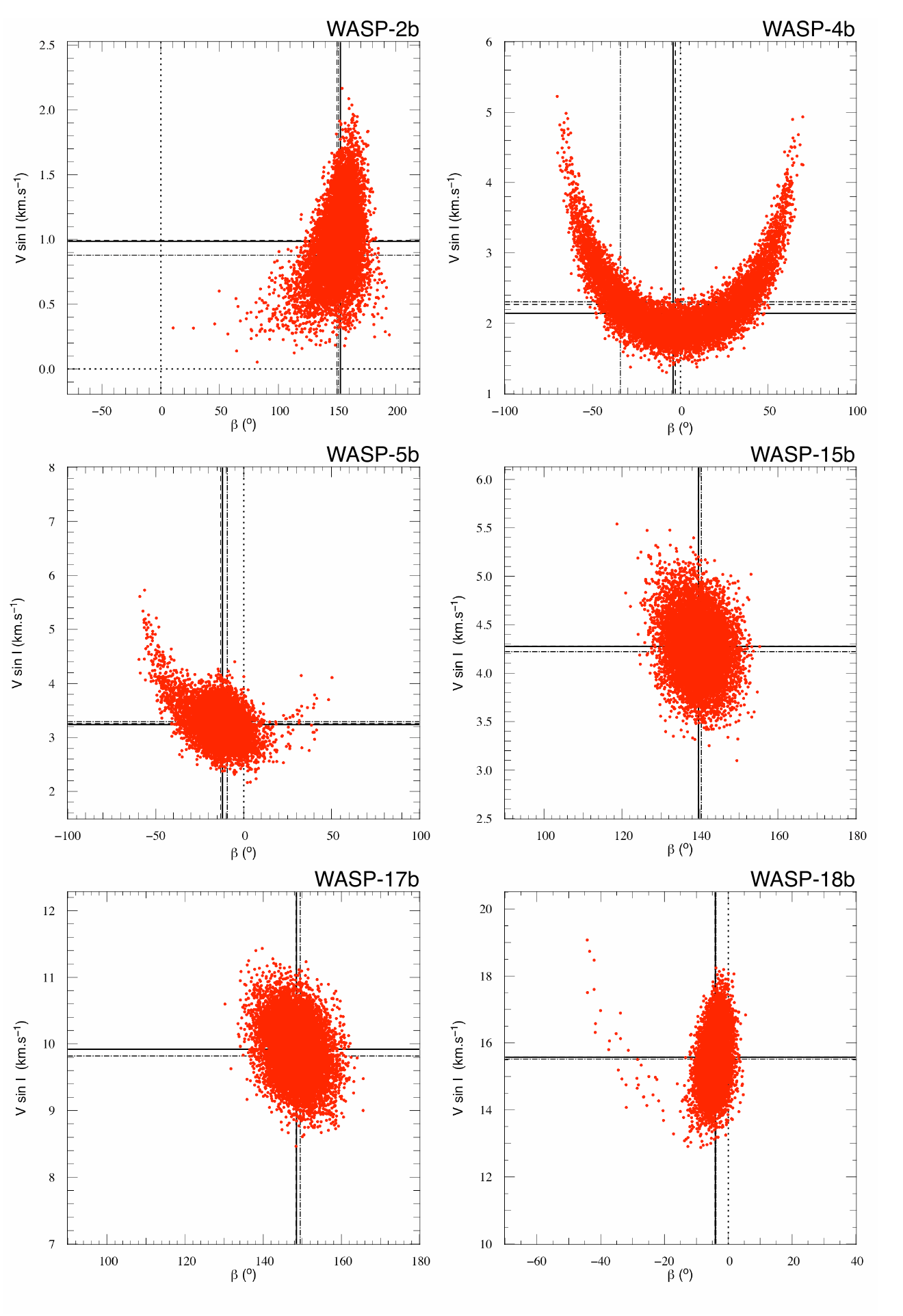}
\caption{Posterior probability distribution issued from the MCMC showing the resulting distributions of points between $V\sin I$ and $\beta$ for our six WASP targets. These distributions are issued from the chains that gave our preferred solutions as explained in the text. The dotted lines show where zeros are, the straight lines represent the medians of the distributions, the dashed lines plot the positions of the average values, the dash-dotted lines indicate the values with the lowest $\chi^2$ (some lines can overlap).  The scale of the boxes was adapted to include the whole distibutions.}\label{fig:betas}

\end{figure*}

Because the impact parameter is small, a degeneracy between $\beta$ and $V\sin I$ appeared, as expected (see Figs.~\ref{fig:WASP4dis}d \& \ref{fig:betas}). The values on stellar rotation for our unconstrained fits reach unphysical values as high as $V\sin I=150$\,km\,s$^{-1}$. We imposed a prior on the stellar rotation to restrict it to values consistent with the spectroscopic analysis. 

The reduced $\chi^2$ is the same within error bars whether eccentricity if fitted or fixed to zero. Therefore the current best solution, by minimising the number of parameters, is a circular orbit. 

The eccentricity is constrained  to $e < 0.0182$. Thanks to the long time series in spectroscopy we also investigated the presence of a long term radial velocity trend. Nothing was significantly detected: $|  \dot{\gamma} | < 30$\,m\,s$^{-1}$\,yr$^{-1}$. 

Because of the small impact parameter the spin-orbit angle is poorly constrained with $\beta= -4^{\circ\,+43}_{\,\,\,\,-34}$, even when a prior is imposed on $V\sin I$. 
The high S/N of the Rossiter-McLaughlin effect allows us to exclude a projected retrograde orbit. 






\begin{sidewaystable*}
\caption{Fitted and physical parameters found after fitting photometric and spectroscopic data of WASP-4b with a prior on $V\sin I$. The value from eccentricity was found using information from other fits as explained in the text.}\label{tab:params}
\begin{tabular}{llllllll}
\hline
\hline
Parameters &$units$ & WASP-2b & WASP-4b & WASP-5b & WASP-15b & WASP-17b & WASP-18b\\
\hline
\\
\multicolumn{2}{l}{\textit{fitted parameters}} &&\\
\\
$D$      &                        &$0.01802^{+0.00027}_{-0.00025}$ &$ 0.023334 ^{+ 0.000044}_{- 0.000072}$         &$ 0.01223 ^{+ 0.00040}_{- 0.00014}$               &$ 0.00969 ^{+ 0.00013}_{- 0.00011}$        &$0.01672^{+0.00020}_{-0.00016}$&$ 0.00916 ^{+ 0.00020}_{- 0.00012}$\\
$K$ &m\,s$^{-1}$         &$153.6^{+3.0}_{-3.1}$                        &$ 242.1 ^{+ 2.8}_{- 3.1}$                                        &$ 268.7 ^{+ 1.8}_{- 1.9}$                                      &$ 64.6 ^{+ 1.20}_{- 1.25}$                            &$52.7^{+3.0}_{-2.8}$                          &$ 1816.7 ^{+ 1.9}_{- 1.9}$\\
$b$ &$R_{\star}$         &$0.737^{+0.012}_{-0.013}$               &$ 0.051 ^{+ 0.023}_{- 0.049}$                              & $ 0.37 ^{+ 0.11}_{- 0.06}$                                   &$ 0.525 ^{+ 0.037}_{- 0.028}$                      &$0.400^{+0.043}_{-0.040}$               &$ 0.527 ^{+ 0.052}_{- 0.046}$ \\
$W$ &days                   &$0.07372^{+0.00065}_{-0.00068}$ &$ 0.08868 ^{+ 0.00008}_{- 0.000014 }$              & $ 0.0988 ^{+ 0.0013}_{- 0.0005 }$                   &$ 0.1547 ^{+ 0.0012}_{- 0.0009}$              &$0.1843^{+0.0013}_{-0.0010}$        &$ 0.09089 ^{+ 0.00080}_{- 0.00061}$\\
$P$ &days                    &$2.1522254^{+0.0000015}_{-0.0000014}$& $ 1.3382299 ^{+ 0.0000023}_{- 0.0000021}$ & $ 1.6284229 ^{+ 0.0000043}_{- 0.0000039}$&$ 3.752100 ^{+ 0.000009}_{- 0.000011}$&$3.7354330^{+0.0000076}_{-0.0000075}$&$ 0.94145290 ^{+ 0.00000078}_{- 0.00000086}$\\
$T_0$ &bjd                   &$3991.51428^{+0.00020}_{-0.00021}$&$ 4387.327787 ^{+ 0.000040}_{- 0.000039}$  & $ 4373.99601 ^{+ 0.00014}_{- 0.00016}$  &$ 4584.69819 ^{+ 0.00021}_{- 0.00020}$ &$4559.18096^{+0.00025}_{-0.00021}$&$ 4664.90531 ^{+ 0.00016}_{- 0.00017}$\\
$e \cos \omega$&     &-& -     &  -    &-&           -  &     $ -0.00030 ^{+ 0.00071}_{- 0.00063}$\\
$e \sin \omega$&      &-& -     &  -     &-&         -    &  $ -0.00845 ^{+ 0.00092}_{- 0.00087}$  \\
\multicolumn{2}{l}{$V\sin I\,\cos \beta$}&$-0.86^{+0.30}_{-0.32}$                  &$ 1.88 ^{+ 0.18}_{- 0.16}$                                     &$ 3.10 ^{+ 0.40}_{- 0.20}$                                   &$ -3.24 ^{+ 0.27}_{- 0.31}$                            &$-8.43^{+0.45}_{-0.52}$& $ 15.52 ^{+ 1.03}_{- 0.67}$\\
\multicolumn{2}{l}{$V\sin I\,\sin \beta$}&$0.44^{+0.17}_{-0.18}$                    &$ -0.13 ^{+ 1.65}_{- 1.33}$                                   & $ -0.68^{+ 0.54}_{- 0.48}$                                  &$ 2.75 ^{+ 0.34}_{- 0.36}$                              &$5.17^{+0.75}_{-0.81}$&$ -1.10 ^{+ 0.68}_{- 0.64}$ \\

\\
\\
\multicolumn{2}{l}{\textit{derived parameters}}       &&      \\
\\
$R_\mathrm{p} / R_{\star}$        &&$0.1342^{+0.0010}_{-0.0009}$  & $ 0.15275 ^{+ 0.00014}_{- 0.00024} $  &$ 0.1106 ^{+ 0.0018}_{- 0.0006} $   &$ 0.09842 ^{+ 0.00067}_{- 0.00058}$&$0.12929^{+0.00077}_{-0.00061}$&$ 0.09576 ^{+ 0.00105}_{- 0.00063}$ \\
\\
$R_{\star} / a$                               &&$0.1248^{+0.0025}_{-0.0024}$  & $ 0.18079^{+ 0.00037 }_{- 0.00040} $  &$ 0.182^{+ 0.011 }_{- 0.004} $   &$ 0.1342 ^{+ 0.0039}_{- 0.0027}$       &$0.1467^{+0.0033}_{-0.0025}$  &$ 0.313 ^{+ 0.012}_{- 0.009}$\\
$\rho_{\star}$   &$\rho_{\odot}$ &$1.491^{+0.088}_{-0.85}$           & $ 1.2667^{+ 0.0084 }_{- 0.0077} $         &$ 0.84^{+ 0.06 }_{- 0.14} $          &$ 0.394 ^{+ 0.024}_{- 0.032}$              &$0.304^{+0.016}_{-0.020}$         &$ 0.493 ^{+ 0.043}_{- 0.051}$\\
$R_{\star}$  &R$_{\odot}$          &$0.825^{+0.042}_{-0.040}$         & $0.903^{+0.016}_{-0.019}$                     &$1.060^{+0.076}_{-0.028}$        &$ 1.440 ^{+ 0.064}_{- 0.057}$              &$1.579^{+0.067}_{-0.060}$         &$ 1.360 ^{+ 0.055}_{- 0.041}$\\
$M_{\star} $  &M$_{\odot}$        &$0.84^{+0.11}_{-0.12}$                & $ 0.930^{+ 0.054 }_{- 0.053}$                  &$ 1.000^{+ 0.063 }_{- 0.064}$   &$ 1.18 ^{+ 0.14}_{- 0.12}$                     &$1.20^{+0.12}_{-0.12}$                &$ 1.24 ^{+ 0.04}_{- 0.04}$\\
$V\sin I$ &km\,s$^{-1}$           &$0.99^{+0.27}_{-0.32}$                 &$ 2.14 ^{+ 0.38}_{- 0.35}$                          &$3.24^{+0.35}_{-0.27}$               &$ 4.27^{+0.26}_{-0.36}$                        &$9.92^{+0.40}_{-0.45}$              &$ 14.67^{+0.81}_{-0.57}$ *\\
\\
$R_\mathrm{p} / a$                               &&$0.01675^{+0.00045}_{-0.00040}$ & $ 0.027617^{+ 0.000064 }_{- 0.000083} $  &$ 0.0201^{+ 0.0015 }_{- 0.0006} $           &$ 0.01321 ^{+ 0.00047}_{- 0.00030}$&$0.01897^{+0.00051}_{-0.00040}$  &$ 0.0299 ^{+ 0.0016}_{- 0.0010}$\\
$R_\mathrm{p}$  &R$_\mathrm {J}$ &$1.077^{+0.055}_{-0.058}$  & $ 1.341^{+ 0.023}_{- 0.029} $                        & $ 1.14^{+ 0.10}_{- 0.04} $                         &$ 1.379 ^{+ 0.067}_{- 0.058}$              &$1.986^{+0.089}_{-0.074}$                & $ 1.267 ^{+ 0.062}_{- 0.045}$\\
$M_\mathrm{p}$ &M$_\mathrm{J}$  &$0.866^{+0.076}_{-0.084}$  & $1.250 ^{+ 0.050}_{- 0.051}$                         &  $1.555 ^{+ 0.066}_{- 0.072}$                  &$ 0.551 ^{+ 0.041}_{- 0.038}$              & $0.453^{+0.043}_{-0.035}$             &$ 10.11 ^{+ 0.24}_{- 0.21}$\\
\\
$a$ &AU                             &$0.0307^{+0.0013}_{-0.0015}$ & $ 0.02320 ^{+ 0.00044}_{- 0.00045} $& $ 0.02709 ^{+ 0.00056}_{- 0.00059} $    &$ 0.0499 ^{+ 0.0019}_{- 0.0017}$       &$0.0500^{+0.0017}_{-0.0017}$  &$ 0.02020 ^{+ 0.00024}_{- 0.00021}$\\
$i$   &$^{\circ}$                 &$84.73^{+0.18}_{-0.19}$             & $ 89.47 ^{+ 0.51}_{- 0.24} $                   &$ 86.1 ^{+ 0.7}_{- 1.5} $                              &$ 85.96 ^{+ 0.29}_{- 0.41}$                   &$86.63^{+0.39}_{-0.45}$              &$ 80.6 ^{+ 1.1}_{- 1.3}$\\
$e$                                &   &$< 0.070$                                        &   $< 0.0182$                                              &$< 0.0351$                                                   &$< 0.087$                                                  &$< 0.110$                                       &$ 0.00848^{+0.00085}_{-0.00095}$\\
$\omega$  &$^{\circ}$     &-                                                        & -                                                                    &-                                                                       &-                                                                  &-                                                         &$ -92.1^{+4.9}_{-4.3}$\\
$\beta$ &$^{\circ}$           &$153^{+11}_{-15}$                       & $-4^{+43}_{-34}$                                      &  $-12.1^{+10.0}_{-8.0}$                             &$ 139.6^{+5.2}_{-4.3}$                            &  $148.5^{+5.1}_{-4.2}$                &$ -4.0^{+5.0}_{-5.0}$\\
 $| \dot{\gamma} |$ &$(m\,s^{-1}\,yr^{-1})$ &$ < 36$              & $ < 30$                                                        & $< 47$                                                           &$ < 11 $                                                     &$< 18$                                             &$< 43$\\
\\
O-C$_\mathrm{\,RV}$  &m\,s$^{-1}$ & 11.73 & 15.16 &12.63&10.89&31.32&13.70\\
O-C$_\mathrm{\,RM}$  &m\,s$^{-1}$ &6.71&6.71&7.72&7.35&31.87; 30.57&15.02\\
\hline

\end{tabular}
\note{* this value is not really a $V\sin I$ but more an amplitude parameter for fitting the Rossiter-McLaughlin effect. Please refer to section \ref{subsec:WASP18} treating WASP-18b}
\end{sidewaystable*}

\begin{figure*}
\centering                     
\includegraphics{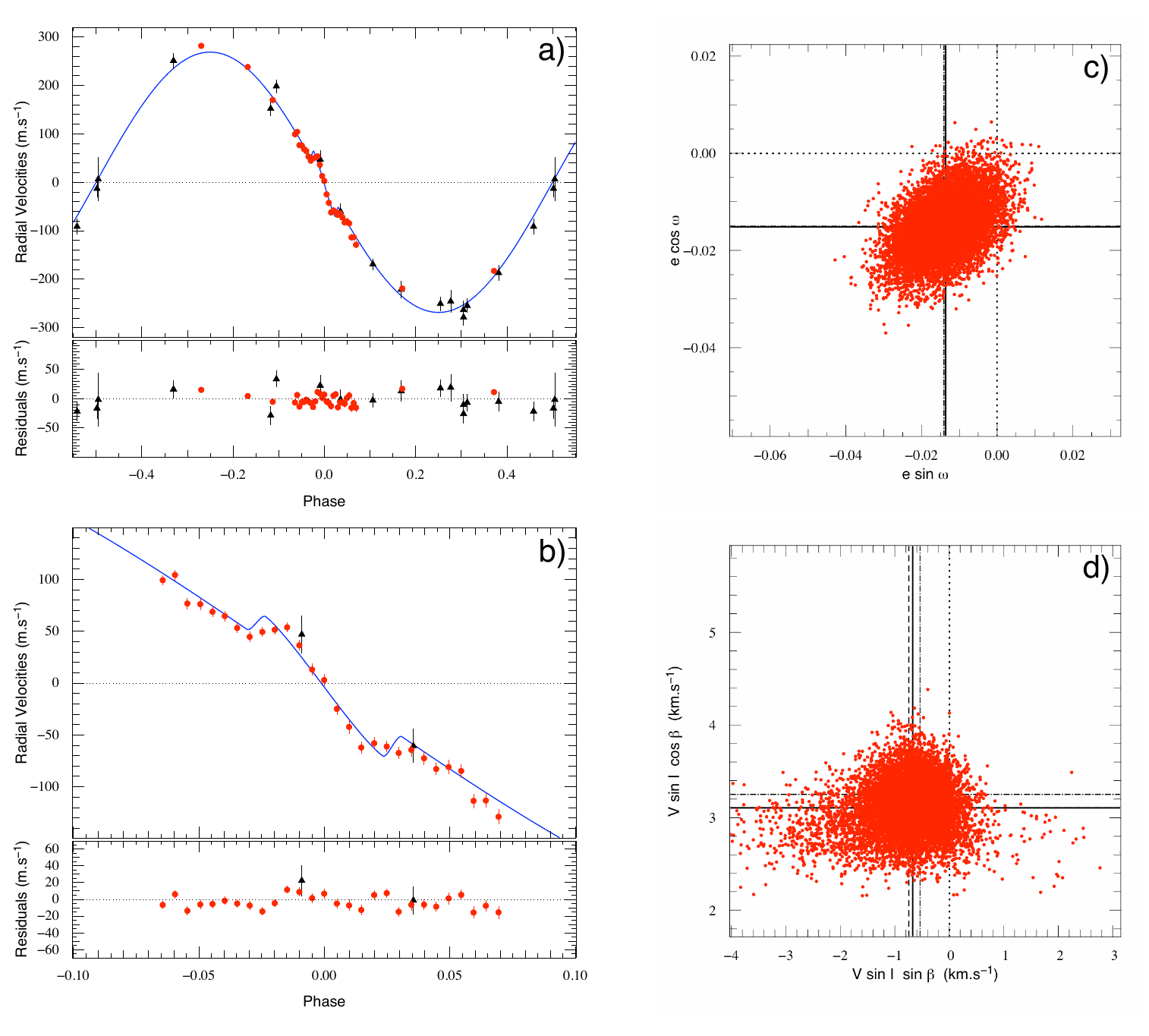}
\caption{
Fit results for WASP-5b. 
\textit{Nota Bene}: Legend similar to the legend in Fig.\ref{fig:WASP2dis}. 
}\label{fig:WASP5dis}
\end{figure*}

\subsection{WASP-5b}\label{subsec:WASP5}


Using HARPS, we took
a series of 28 exposures 
on WASP-5 at a cadence of roughly 630s with a mean photon noise of $5.5$\,m\,s$^{-1}$ on 2008 October 16. Other measurements were obtained at dates before and after this transit. 
Five additional CORALIE spectra were acquired the month before the HARPS observations. They were taken about a year after the data published in \citet{Anderson:2008p1379}.  All spectroscopic data is available from the appendices.

To help determine transit parameters, published photometry was assembled and comprises three seasons of WASP data, two \textit{C2 Euler} lightcurves in $R$ band, and one \textit{FTS} $i'$ band lightcurve \citep{Anderson:2008p1379}.

WASP-5b's 49 RV measurements and 14\,754 photometric points were fitted with up to 10 free parameters to which 8 adjustment parameters had to be added: two $\gamma$ velocities and six normalisation factors. 

\begin{figure*}
\centering                     
\includegraphics{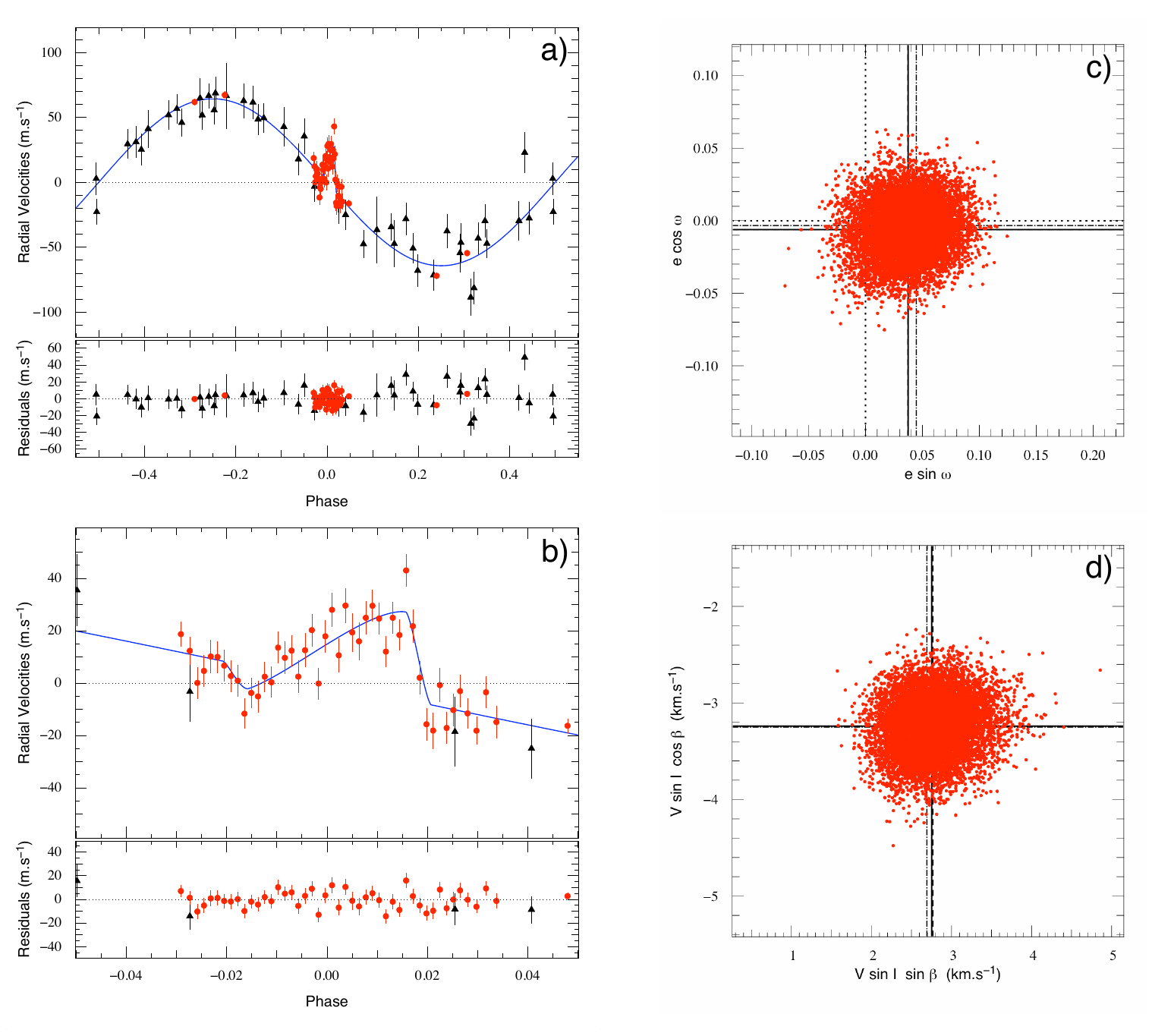}
\caption{
Fit results for WASP-15b. 
\textit{Nota Bene}: Legend similar to the legend in Fig.\ref{fig:WASP2dis}. 
}\label{fig:WASP15dis}
\end{figure*}


The imposition of a prior on $V\sin I$ prior makes little impact on the final results (see appendices) and their $1\sigma$ error bars but prevents $V\sin I$ to go to unphysical values when, through the random process of the MCMC, the impact parameter $b$ gets very close to zero on a few occasions. We choose the solution using a prior.
The priorless solution gives a $V\sin I$ fully consistent with \vsini thereby obtaining an independent measurement of the projected stellar equatorial rotation speed. 
Allowing eccentricity to float did not produce a significantly better fit. It has a 99.6\,\% chance of being different from zero: at $2.9\,\sigma$. Thus, minimising the number of parameters for a similar fit,
 we chose the solution with a circular orbit and simply place an upper limit on the eccentricity: $e < 0.0351$. No long term RV trend appears at this date: $| \dot{\gamma} | < 47$\,m\,s$^{-1}$\,yr$^{-1}$.

Parameters extracted are similar to those that were published in  \citet{Gillon:2009p1630} \& \citet{Anderson:2008p1379} and with \citet{Southworth:2009p6564} using a independent dataset. The projection of the spin-orbit angle is found to be: $\beta = -12.1^{\circ\,+10.0}_{\,\,\,\,\,-8.0}$ and we obtain an independent measurement of $V\sin I = 3.24^{+0.34}_{-0.35}$\,km\,s$^{-1}$ fully compatible with the spectral value that was used as a prior in other fits. Results are presented in Table \ref{tab:params}.

The $\chi^2_\mathrm{reduced }$ for spectroscopy (see Table \ref{tab:WASP1comp}) is quite large, at $3.68\pm0.44$. The $O-C$ for CORALIE data stand at 17.94\,m\,s$^{-1}$ to be compared with an average error bar of 18.13\,m\,s$^{-1}$. The badness of fit therefore comes from the HARPS sequence which has a dispersion of 8.98\,m\,s$^{-1}$ for an average error bar of 5.49\,m\,s$^{-1}$. From Fig.~\ref{fig:WASP5dis}b we can see that residuals are quite important during the transit; Fig.~\ref{fig:WASP5dis}d also shows that the MCMC does not produce a clean posterior distribution. This is mostly caused by impact parameter values nearing zero during parameter exploration and causing a degeneracy between $V\,\sin\,I$ \& $\beta$. This can be observed on Figure \ref{fig:betas} with similitude to what occurs to WASP-4.

No better solution can be adjusted to the data: we remind that the RM effect is fitted in combination with six photometric sets which strongly constrain the impact parameter, depth and width of the Rossiter-McLaughlin effect. The $V\sin I\,\cos \beta$ vs  $V\sin I\,\sin \beta$ distribution is not centred on zero but close to it. This may come from the intrinsic dispersion in the data. Among the six data points which are spread over the rest of the phase, we have a dispersion of 11.92\,m\,s$^{-1}$. A likely cause to explain the data dispersion is stellar activity. Table~\ref{tab:SpecParam} indicates that this star is moderately active. A longer discussion on $\chi^2_\mathrm{reduced} > 1.$ is presented in section~\ref{subsec:chi2}. \citet{Santos:2000p6686} show that for the $\log R^\prime_\mathrm{HK}$ that we find, we can expect a variation in velocities of the order of 7 to 12\,m\,s$^{-1}$. 

\begin{figure*}
\centering                     
\includegraphics{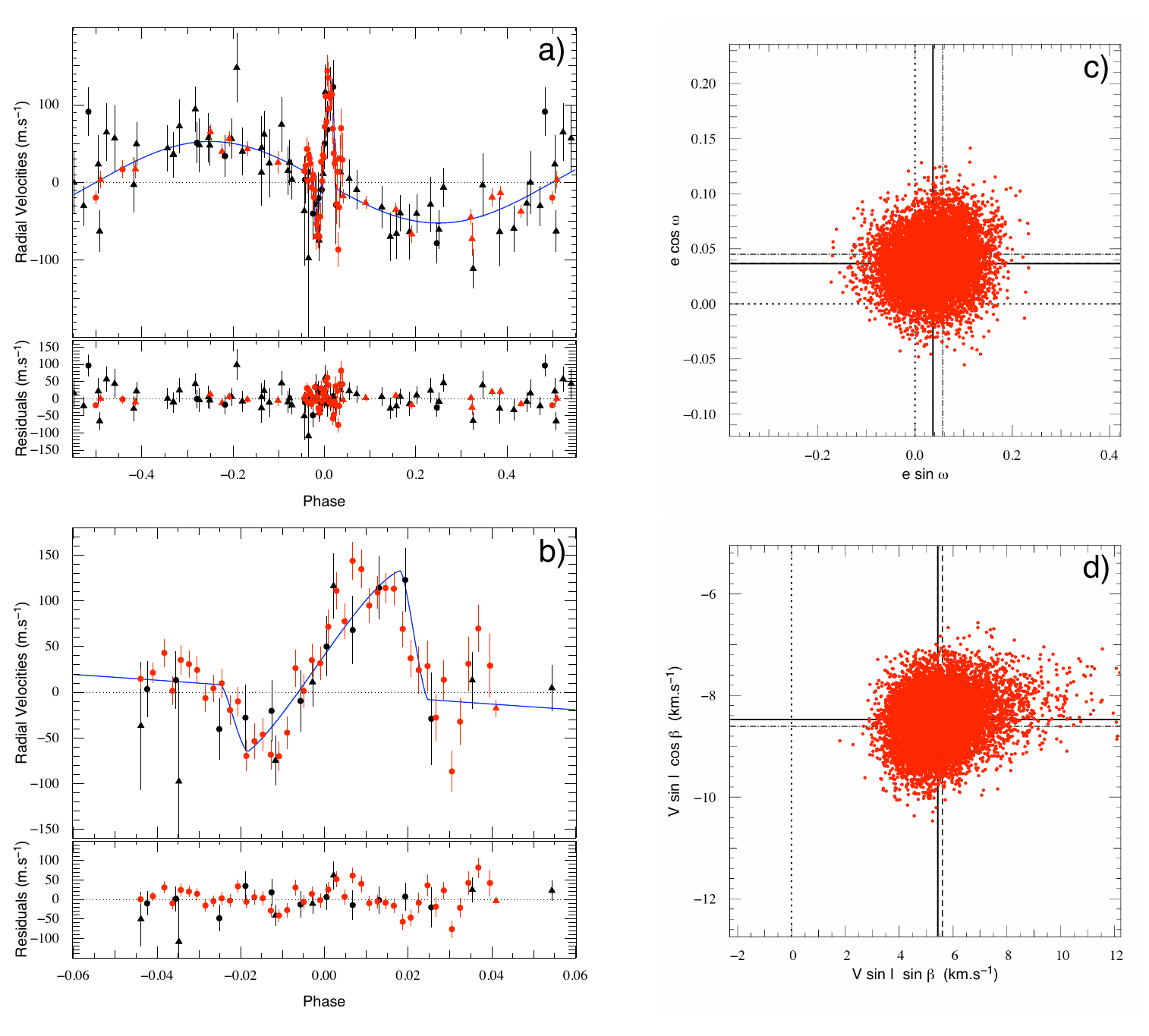}
\caption{
Fit results for WASP-17b. 
On a) and b) black circles represent the RM effect taken with CORALIE, while  black triangles picture the remaining CORALIE measurements;  red dots show the HARPS RM data, red triangles are the remained HARPS points. 
\textit{Nota Bene}: Legend similar to the legend in Fig.\ref{fig:WASP2dis}. 
}\label{fig:WASP17dis}
\end{figure*}

\subsection{WASP-15b}\label{subsec:WASP15}

Observations were conducted using the spectrographs CORALIE and HARPS. 23 new spectra have been acquired with CORALIE in addition to the 21 presented in \citet{West:2009p2783} and extending the time series from about a year to 500 days. 
We observed  a transit with HARPS on 2009 April 27. 46 spectra were obtained that night, 32 of which are during transits with a cadence of 430s. Additional observations have been taken as noted in the journal of observations.

The photometric sample used for fitting the transit has data from five time-series
in the WASP bandpass, as well as one $I$ and one $R$ band transit from \textit{C2 Euler} \citep{West:2009p2783}. The spectral data were partitioned into two sets: CORALIE and HARPS. 

7 normalisation factors and 2 $\gamma$ velocities were added to ten free floating parameters to adjust our models to the data which included a total of 95 spectroscopic observations and 23\,089 photometric measurements.



For the various solutions attempted,  $\chi^2_\mathrm{reduced}$ are found the same (Table \ref{tab:WASP2comp}). 
We therefore choose the priorless, circular adjustment as our solution.

Compared to \citet{West:2009p2783}, parameters have only changed little. Thanks to the higher number of points we give an upper limit on eccentricity: $e < 0.087$ (Fig.~\ref{fig:WASP15dis}c shows results consistent with zero); there is no evident long term evolution in the radial velocities, which is constrained within: $| \dot{\gamma} | < 11$\,m\,s$^{-1}$\,yr$^{-1}$.  
The projected spin-orbit angle is found rather large with $\beta = 139.6^{\circ\,+5.2}_{\,\,\,\,-4.3}$ making WASP-15b appear as a retrograde planet with a very clear detection.  $V\sin I$ is found within $1\,\sigma$ of the spectrally analysed value of \vsini from \citet{West:2009p2783} at $4.27^{+0.26}_{-0.36}$\, km\,s$^{-1}$ and as such constitutes a precise independent measurement. 

$\chi^2_ \mathrm{reduced}=1.51\pm0.19$ for the spectroscopy, indicating a good fit of the Keplerian as well as of the Rossiter-McLaughlin effect, the best fit in this paper. Full results can be seen in Table \ref{tab:params}.





\begin{figure*}
\centering                     
\includegraphics{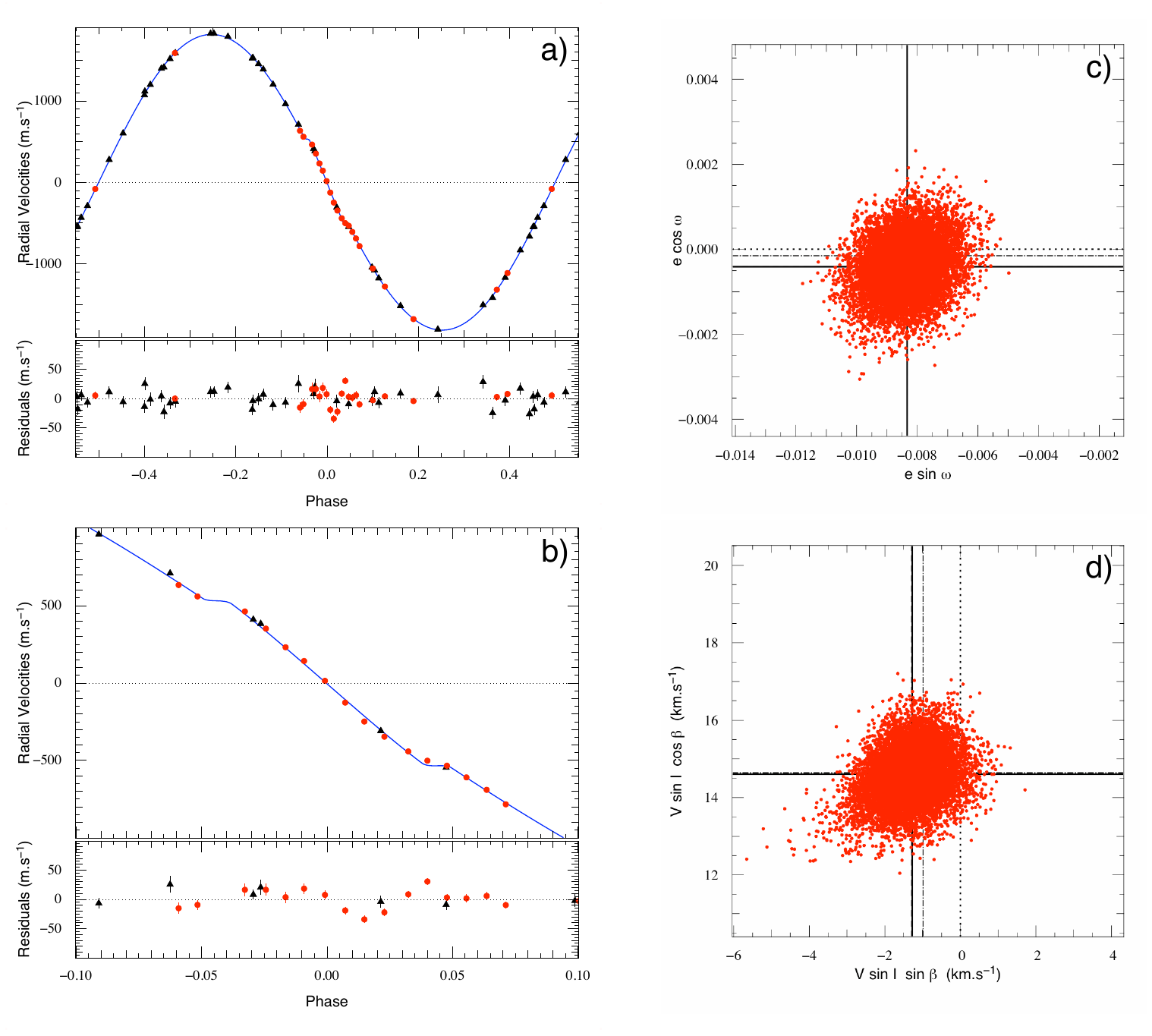}
\caption{
Fit results for WASP-18b. 
\textit{Nota Bene}: Legend similar to the legend in Fig.\ref{fig:WASP2dis}. 
}\label{fig:WASP18dis}
\end{figure*}

\subsection{WASP-17b}\label{subsec:WASP17}


On 2009 May 22, 11 CORALIE spectra were obtained at a cadence of 2030s with an average precision of $33.67$\,m\,s$^{-1}$ to confirm the detection of retrograde orbital motion announced by \citet{Anderson:2010p5177}. The sequence was stopped when airmass reached 2. HARPS was subsequently used and on 2009 July 5 a sequence of 42 spectra was acquired with a cadence of 630s during transit. They have a mean precision of $19.02$\,m\,s$^{-1}$. 
In addition to these and to data already published 12 CORALIE spectra and 15 HARPS spectra were obtained. All the spectroscopic data is presented in the appendices.

The photometry includes five timeseries of data in the WASP bandpass, and one $C2$ $Euler$ $I$ band transit \citep{Anderson:2010p5177}.

The model had to adjust up to 10 free floating parameters and 10 adjustment parameters (6 photometric normalisation factors and 4 radial velocity offsets) to 15\,690 photometric data points and 124 spectroscopic points. 



\begin{table}
\caption{List of $\gamma$ velocities for WASP-17's RV sets.}\label{tab:gamma}
\begin{tabular}{lll}
\hline
\hline
Instrument & Dataset & $\gamma$ (m\,s$^{-1}$)\\
\hline
\\
CORALIE & \textit{Rossiter-McLaughlin effect} & $-49500.80^{+2.62}_{-1.57}$\\
CORALIE & \textit{orbital Doppler shift} & $-49513.67^{+0.46}_{-0.37}$\\
HARPS & \textit{Rossiter-McLaughlin effect} & $-49490.59^{+2.72}_{-1.64}$\\
HARPS & \textit{orbital Doppler shift} & $-49491.68^{+0.17}_{-0.17}$\\
\\
\hline

\end{tabular}
\end{table}

The RV was separated into four datasets fitted separately as detailed in Table \ref{tab:gamma}. 
This was done to mitigate the possibility that the RM effect was observed at a particular activity level for the star. Stellar activity adds an additional RV variation. For a set where this data is taken randomly over some time, one expects activity to act like a random scatter around a mean which would be the true $\gamma$ velocity of the star in space. But for a sequence such as the RM effect, we expect only a slowly-varying radial-velocity bias caused by the activity level on the star on the night concerned. This  analysis method is explained in \citet{Triaud:2009p3865} which showed an offset in $\gamma$ velocities between different Rossiter-McLaughlin sequences of HD\,189733 which can only be attributed to stellar variability. The large number of CORALIE and HARPS measurements outside transit and their large temporal span allowed us to separate RV sets for WASP-17 but not for the other targets. Table \ref{tab:gamma} shows the four values of $\gamma$. We remark a difference of 13\,m\,s$^{-1}$ for CORALIE, justifying our segmentation of the data.

Among the four computed chains, we select the circular solution, with prior on $V\,\sin\,I$ since our results show eccentricity is not significantly detected but that the prior on $V\sin I$ prevents the MCMC from wandering to small impact parameters leading to the degeneracy between $V\,\sin\,I$ and $\beta$.

The non significant eccentricity presented by \citet{Anderson:2010p5177} was not confirmed, so a circular orbit was adopted. 
We confine to within $e<0.110$. Eccentricity affects the derived value of the stellar density, and thereby also affects the planet's radius measurement. Our circular solution suggests that WASP-17b's radius is $1.986^{+0.089}_{-0.074}\, R_\mathrm{J}$, making it the largest and least dense extrasolar planet discovered so far. We looked for an additional long term acceleration but found none: $| \dot{\gamma} | < 18$\,m\,s$^{-1}$\,yr$^{-1}$.

The Rossiter-McLaughlin effect is well fitted. The residuals show some dispersion about the model during the HARPS sequence. At the end of the HARPS transit, the airmass attained high values which account for the larger error bars, the sparser sampling and higher dispersion. By comparison the CORALIE sequence appears better: its longer exposures blurred out short-term variability. Both $V\sin I$ and $\beta$ are unambiguously detected. WASP-17b is on a severely misaligned orbit: $V\sin I = 9.92$\,km\,s$^{-1}$ and $\beta = 148.5^{\circ\,+5.1}_{\,\,\,\,-4.2}$. Full results are displayed in Table~\ref{tab:params}.

\subsection{WASP-18b}\label{subsec:WASP18}


Soon after WASP-18b was confirmed by the spectrograph CORALIE, a Rossiter-McLaughlin effect was observed with HARPS. We obtained 19 measurements at a cadence of 630s on 2008 August 21. The mean photon noise for the transit sequence is $6.99$\,m\,s$^{-1}$. Seeing and airmass improved during the sequence, increasing the S/N and decreasing the individual error bars. Additional data were also acquired out of transit.
\citet{Hellier:2009p3885} presented 9 RV measurements from CORALIE. 28 more have been taken and are presented in this paper. They span over three months. The total data timeseries spans close to 500 days. All RV measurements are presented in the journal of observations at the end of the paper.

Transit timing and geometry were secured by four photometric series: two SuperWASP seasons and two \textit{C2 Euler} transits in $R$ band, presented in \citet{Hellier:2009p3885}.

The fitted data comprises 8593 photometric measurements and 60 radial velocities. Ten free parameters were used, with, in addition, four normalisation constants and two $\gamma$ velocities. 




Eccentricity is clearly detected, improving $\chi^2_\mathrm{reduced}$ from $5.58 \pm 0.47 $ to $3.70 \pm 0.36$ (from $4.31\pm 0.46$ to $2.00\pm0.32$ if we remove the RM effect from the calculation). We therefore exclude a circular solution.

The $V\sin I$ found in the priorless chain differs from the spectral analysis ($15.57^{+1.01}_{-0.69}$ instead of $11\pm 1.5$\,km\,s$^{-1}$), this solution is preferred so as to not produce biased results. For this particular case, we should consider $V\,\sin\,I$ more like a amplitude parameter in order to fit the Rossiter-McLaughlin effect rather than a bone fide measurement of projected rotation of the star. Therefore, the solution we favour is that of an eccentric orbit, without a prior on the $V\sin I$. 

Results are presented in Table.~\ref{tab:params}, and the best fit is shown in Fig.~\ref{fig:WASP18dis}. This Rossiter-McLaughlin effect is one of the largest so far measured, with an amplitude of nearly $185$\,m\,s$^{-1}$. During the transit sequence $O-C$ = 15.02 m\,s$^{-1}$ for a mean precision of 6.95 m\,s$^{-1}$: the fit is poor; $\chi^2_\mathrm{reduced} = 3.70 \pm 0.36$. This is likely caused by a misfit of a symmetric Gaussian on a no longer symmetrical CCF\footnote{this was noted in \citet{Triaud:2009p3865} in the case of HD\,189733b and CoRoT-3b, but can also be seen on fits of CoRoT-2b \citep{Bouchy:2008p229}, Hat-P-2b \citep{Loeillet:2008p881} and others.}. We are in fact \textit{resolving} the planet transit in front of the star like spots can be detected via Doppler tomography. This has recently been modelled and detected for HD\,189733b, as a \textit{Doppler shadow} by \citet{CollierCameron:2010p5889}. 
The accuracy on the $\beta$ parameter is not affected by the misfit since it is measured from the asymmetry of the Rossiter-McLaughlin effect. It is in essence estimated from the difference of time spent between the two hemispheres of the star.


Therefore all parameters can be trusted except the $V\sin I$, including the much sought after $\beta$ angle. We find it to be consistent with zero within $1.5\,\sigma$: $\beta=-4.0^{\circ\,+2.5}_{\,\,\,\,-2.5}$. The precision on this angle is the best we measured, something that is not reflected in the fit, we therefore doubled error bars to $\beta=-4.0^{\circ\,+5.0}_{\,\,\,\,-5.0}$. This is in part thanks to the brightness of the star, allowing precise  measurements of a large amplitude effect. Any departure from the model is quickly penalised in $\chi^2$ by the data. Similarly, eccentricity is detected above $9\,\sigma$ with $e = 0.00848^{+0.00085}_{-0.00095}$ thanks to the large amplitude of the reflex motion. 
Note, that fitting $e\,\cos\,\omega$ and $e\,\sin\,\omega$ can correspond to fitting $e$ proportional to $e$ and tending to bias the search for solutions towards higher values. We attempted a few control fits exploring instead $\sqrt{e} \cos \omega$ \& $\sqrt{e} \sin \omega$. The results showed there is no bias in our analysis, so strongly is the eccentricity constrained by the radial velocities.
The spectroscopic coverage gives us the chance to put some limits on an undetected long term radial velocity drift: $| \dot{\gamma} | < 43$\,m\,s$^{-1}$\,yr$^{-1}$.



The other parameters are consistent with what has been published by Hellier et al. 2009 and are presented in Table~\ref{tab:params}.





\section{Overall results}

Our fits to the Rossiter-McLaughlin effect confirm the presence of planetary
spectroscopic transit signatures in all six systems.  While three of the six appear closely aligned, the other three exhibit highly-inclined, apparently
retrograde orbits. The orbits of all six appear close to circular. Only the massive WASP-18b yields a significant detection of orbital eccentricity.



\subsection{Orbital eccentricities}

As observed in \citet{Gillon:2009p4081}, treating eccentricity as a free fitting parameter increases the error bars on other parameters; we are exploring a larger parameter space. One might argue that allowing eccentricity to float is necessary since no orbit is perfectly circular, therefore making an eccentric orbit the simplest model available. We argue against this for the simple reason that if statistically we cannot make a difference between an eccentric and a circular model then it shows that the eccentric model is not detected. Actually, the mere fact of letting eccentricity float biases the result towards a small non zero number, a bias which can be larger than the actual physical value  \citep{Lucy:1971p5669}. Hence letting eccentricity float when it is not detected  is to allow values of parameter space for all parameters to be explored which do not need to be. This is why, unless $\chi^2$ is significantly improved by adding two additional parameters to a circular model, we consider the former as preferable. 
In addition to the risk of biasing, there is a strong assumption that due to tidal effects circularising the planet's orbit, eccentricities are really small and therefore undetectable for the majority of targets. It is therefore reasonable to assume a value of zero when the data does not contradict it.
To facilitate comparison, we also present the results of fits with floating eccentricity. These are given in the appendices; our preferred solutions are described in the text and in table \ref{tab:params}.

Only for WASP-18b, have we detected some eccentricity in the orbit, thanks primarily to
the high amplitude of the RV signal and the brightness of the target. 
The amount of RV data taken on WASP-18b is not really more than for the other targets. In addition to a high semi-amplitude, sampling is another key to fixing eccentricity properly. The lack of measured eccentricities on our other targets shows how difficult it is to measure a small eccentricity for these planets as long as no secondary transit is detected to constrain it. 
Spurious eccentricities tend to appear in fits to data sets where the
radial velocities are not sampled uniformly around the orbit, and where
the amplitude is small compared to the stellar and instrumental noise levels.

A good example is the case of WASP-17b for which the doubling of high precision RV points solely permitted us to place a tighter constraint compared to \citet{Anderson:2010p5177}.


\subsection{Fitting the Rossiter-McLaughlin effect}\label{sec:psi}

Our observations yielded results from which five sky-projected spin-orbit angles $\beta$ have been determined with precision better than $15^{\circ}$. Three of these angles appear to be retrograde: half our sample. Adding the two other stars from our original sample that have been published separately (WASP-6b and WASP-8b) we obtain 4 out of 8 angles being not just misaligned but also over $90^{\circ}$. The precision on the angle depends mostly on the spectroscopy as is shown by comparing fits where parameters controlled by the photometry are kept fixed (in the appendices).

The error bar on WASP-4b's $\beta$ is large. A degeneracy appears when the impact parameter is close to 0 between $V\sin I$ and $\beta$. The estimate of the spin-orbit angle therefore relies on a good estimate of the stellar rotational velocity as well as with getting a stronger constraint on the shape of the Rossiter-McLaughlin effect. WASP-5b, WASP-17b and WASP-18b are also affected by this degeneracy, with much lower consequences, when the MCMC takes a random step in low impact parameters. This is controlled by the use of a prior on  $V\sin I$.

When the planet is large compared to the parent star, or the star rotates rapidly, the cross-correlation function
develops a significant asymmetry during transit. This happens because the spectral signature of the light blocked by the planet is partially resolved.
Fitting a Gaussian to such a profile yields a velocity estimate that differs systematically from the velocity of the true light centroid. 
\citet{Winn:2005p83}, and later \citet{Triaud:2009p3865} and \citet{Hirano:2010p5304} showed how this effect
can lead to over-estimation of $V\sin I$. \citet{Hirano:2010p5304} have developed an analytic method to compensate for this bias. \citet{CollierCameron:2010p5889} circumvent the problem altogether by modelling the CCF directly, decomposing the profile into a stellar rotation profile and a model of the light blocked by the planet.


Only one star in our sample suffers from this misfit: WASP-18b where easily we see that the value the fit issues for the $V\sin I$ is above the estimated value taken via spectral analysis. WASP-17b is the second fastest rotating star. If affected, it is not by much: the fitted $V\sin I$ is found within $1\,\sigma$ of the \vsini.


\begin{figure}
\centering                     
\includegraphics[width=9cm]{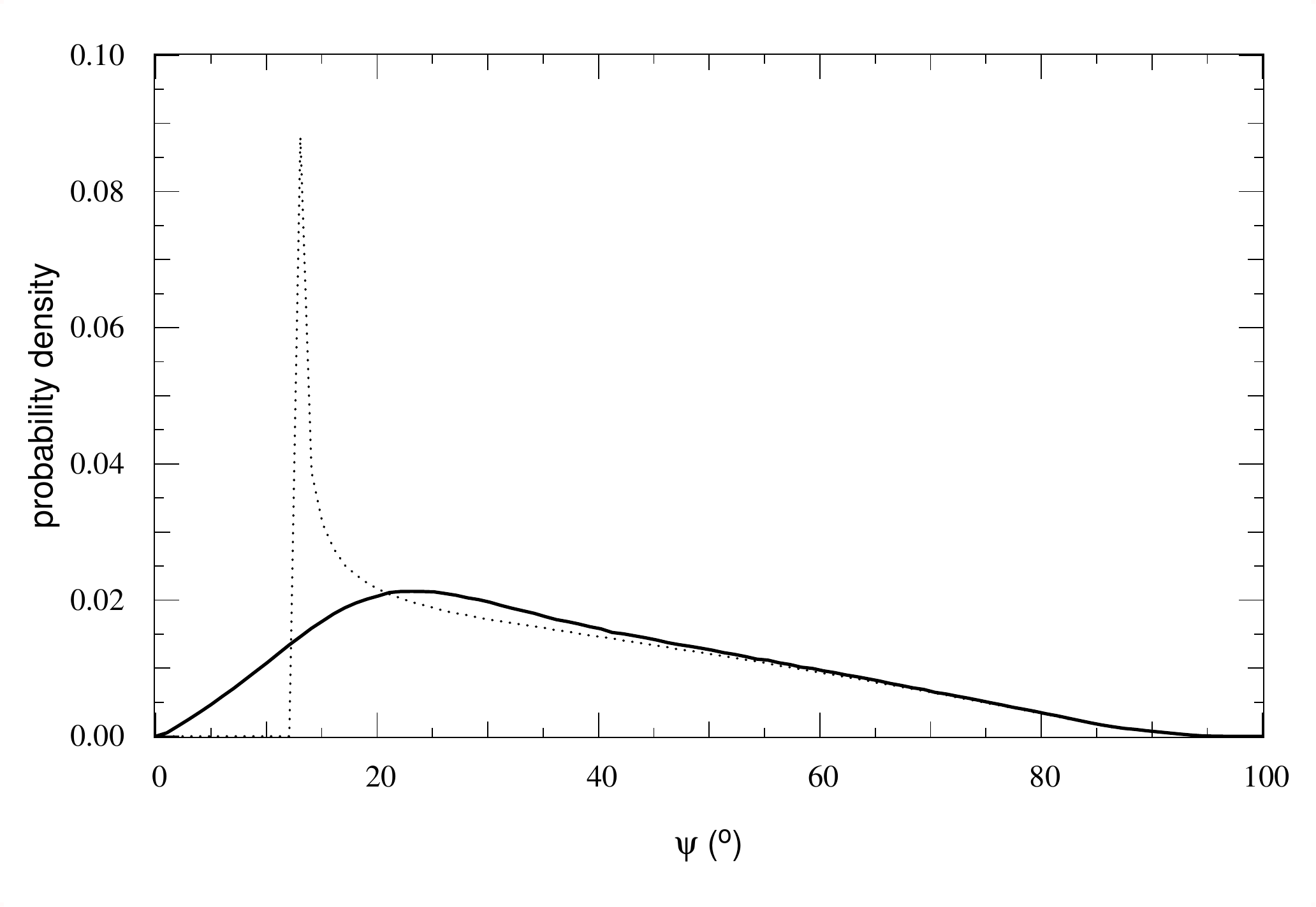}
\includegraphics[width=9cm]{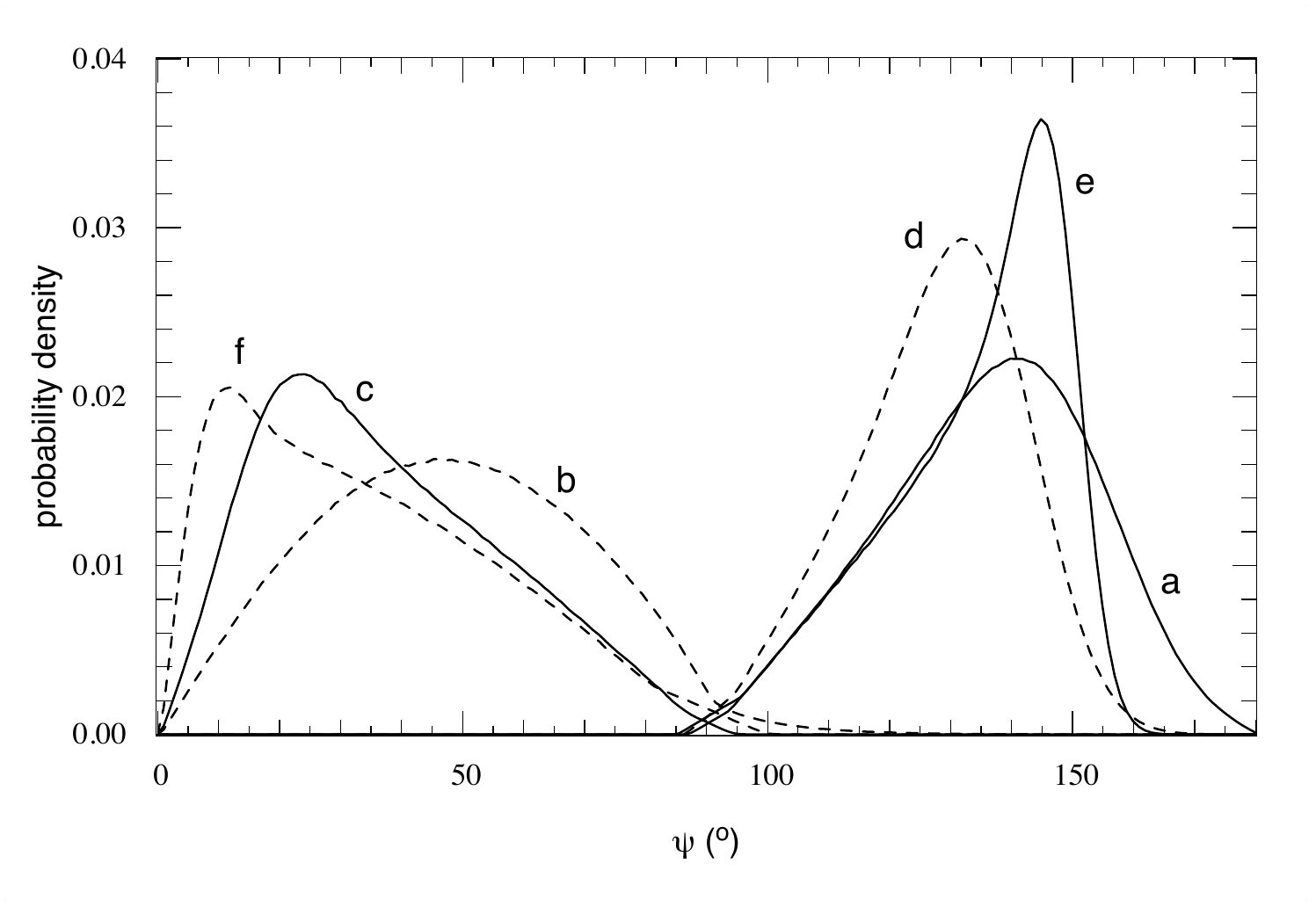}
\caption{\textit{top} Smoothed histogram of the $\psi$ distribution for WASP-5b. The dotted line is when errors on $i$ and $\beta$ are set to zero. The plain curve shows the same conversion from $\beta$ to $\psi$ but with all errors accounted for. \textit{bottom}: 6 smoothed histograms of the distribution in $\psi$ our six targets: a)-WASP-2b  b)-WASP-4b  c)-WASP-5b  d)-WASP-15b   e)-WASP-17b  f)-WASP-18b. Bins are of $1^{\circ}$.}\label{fig:psi_dist}

\end{figure}

\bigskip

 


As shown in \citet{Fabrycky:2009p1845}, we can get an idea of the real angle $\psi$ from $\beta$ by using the following equation, coming only from the geometry of the system:

\begin{equation}\label{eq:psi}
\cos \psi = \cos I \cos i + \sin I \sin i \cos \beta
\end{equation}
where $I$ is the inclination of the stellar spin axis and $i$ the inclination of the planet's orbital axis to the line of sight.

Using the reasonable assumption that the stellar spin axis angle $I$ is distributed isotropically, we computed the above equation using a simple Monte-Carlo simulation to draw a random uniform distribution in $\cos I$. We also inserted the error bars on $i$ and $\beta$, using a Gaussian random number adjusted to the $1\,\sigma$ error bars printed in table \ref{tab:comp}. Fig. \ref{fig:psi_dist} shows the transformation from $\beta$ to $\psi$ for our targets, also illustrating the importance of including error bars in the calculation. We computed the lower $\psi$ (at the $3\,\sigma$ limit) and found that in the stars we surveyed: WASP-15b is $> 90.3^{\circ}$ and WASP-17b $> 91.7^{\circ}$ therefore retrograde, while  WASP-2b is $> 89.8^{\circ}$ most probably retrograde.


Statistically we will fail to detect a Rossiter-McLaughlin effect (hence $\beta$ and $\psi$) on stars nearly pole-on (with a low $I$).
WASP-2b, with its small $V\sin I$ could be a close case. It could be one reason why its RM amplitude is so small (or stellar rotation so low). We observe that the spread in $\psi$ is larger than for our other targets. 

\subsection{$\chi^2_\mathrm{reduced} > 1.$}\label{subsec:chi2}

It can be remarked from the text or from the appendices that a few of our objects have $\chi^2_\mathrm{reduced} > 1.$; in the case of WASP-5 notably. This shows the models are not adjusted perfectly to the data. As showcased by model fits to WASP-4, 15 and 17 (with $\chi^2_\mathrm{reduced} < 2.$) and in many publications using the CORALIE and HARPS spectrographs,  produced error bars on individual radial velocity points are well estimated and understood and worth using as they are \citep{Lovis:2006p6893}. 

An easy way to solve the problem would be to scale error bars so as to achieve an acceptable $\chi^2$. But increasing error bars is not making the model more just; rather, it hides that we do not understand everything: that an extra signal is observed by the instruments; there is information in a bad $\chi^2$. Error bars can be scaled with a value of \textit{stellar jitter} added quadratically to individual errors, but this applies only if one samples randomly over long periods of time an extra stellar signal. In our case, part of our out-of-transit RVs would feel this \textit{jitter}, but it would not apply in the same way to the Rossiter-McLaughlin sequences during which we are sensitive to a correlated noise of a different frequency. This renders the increase of error bars prone to errors of judgement, thus leading to a wrong computation of the model.

Hence, we decided to produce results without interfering with the way the data is estimated giving the best optimisation of the data that we could produce using known and substantiated physics. We leave to the reader the assessment of where this extra signal originates from.

\subsection{Correlations between parameters}

We present a compilation of results from all known observations of the Rossiter-McLaughlin effect  in transiting exoplanetary systems in Table \ref{tab:comp}. No clear correlation is evident between important planetary parameters such as radii, masses, eccentricities, orbital periods, $\beta$ and $V\sin I$, except that planets with $M<\, 2 \, M_\mathrm{J}$ and $e>0.1$ are rare among transiting systems (the only two are Neptunes around M dwarfs); this remark is independent from having a Rossiter-McLaughlin measurement or not.  It is hard to see if this is really a result, or a bias due to observations (eg: transits harder to extract from the survey photometry, or to confirm via radial velocity), or a lack of precision during follow-up making eccentricity hard to detect with confidence. WASP-17b, for example, was previously thought to be the most eccentric transiting planet with $M<\, 2 \, M_\mathrm{J}$ but our analysis yields only an upper limit $e<0.110$. Eccentricities with as great as $e=0.1$ have been published for some planets with masses less that $2 M_\mathrm{J}$, none of these results are significant at more than the $\sim 2\,\sigma$ level.

The current ($M_\mathrm{p} \sin i$, $e$) distribution in radial velocity does not show this result, but these masses are only minimum masses. 

Amongst planets where eccentricity is firmly detected, four out of seven are misaligned. Some of the hot Jupiters appear to be in multiple systems but this appears unrelated to other parameters such as eccentricity or misalignment. Examples are: HD\,80606 \citep{Naef:2001p2419}, HD\,189733 \citep{Bakos:2006p6038}, Hat-P-1 \citep{Bakos:2007p2581}, WASP-8 \citep{Queloz:2010p7085}, Hat-P-7 \citep{Winn:2009p3712}, WASP-2, TrES-2 and TrES-4 \citep{Daemgen:2009p5950}. 



\section{Discussion}


After a long sequence of closely-aligned planets \citep{Fabrycky:2009p1845}, the sudden appearance of so many misaligned planets is somewhat surprising if not unpredicted. 
In a collapsing gas cloud, conservation of angular momentum will create a disc from which a star can form. Thus it is expected that star and disc rotate in the same direction with parallel spin axes. If planets form in and migrate through the disc, we can extend the idea that planets' orbital axes and stellar rotation axes ought to be parallel. Tides alone cannot make a planet retrograde \citep{Hut:1981p2945}. Therefore it is expected that the creation of retrograde planets involves another body: planetary or stellar. Several papers \citep{Wu:2007p4179,Fabrycky:2007p3141, Nagasawa:2008p2997, Chatterjee:2008p2971, Juric:2008p2882, Bate:2000p2978, Bate:2009p3276} produce via various processes, orbits which are not coplanar with the host star's equator. Of these papers \citet{Wu:2007p4179}, \citet{Fabrycky:2007p3141} and \citet{Nagasawa:2008p2997} produce the largest range of angles.


\begin{figure}
\centering                     
\includegraphics[width=9cm]{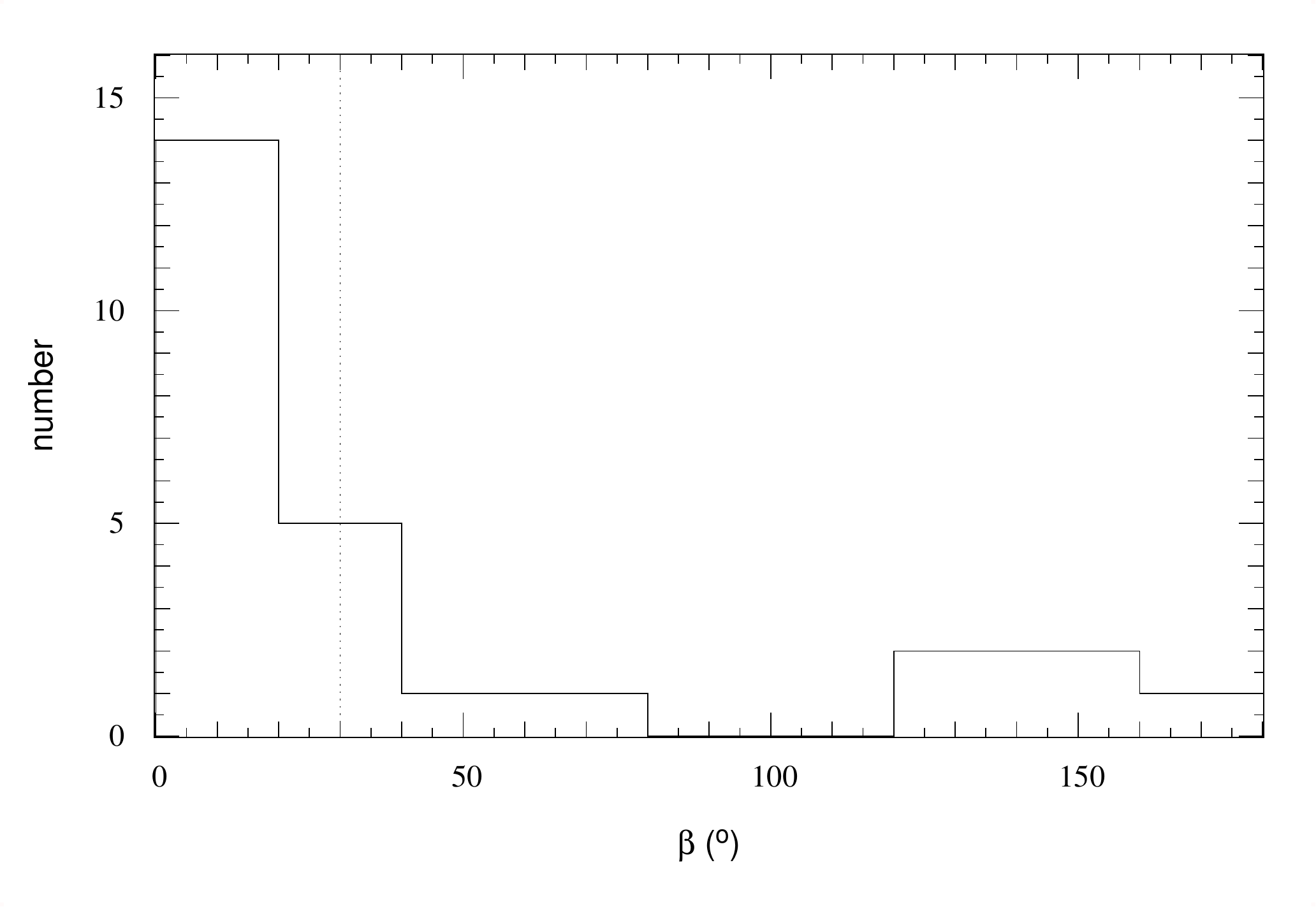}
\includegraphics[width=9cm]{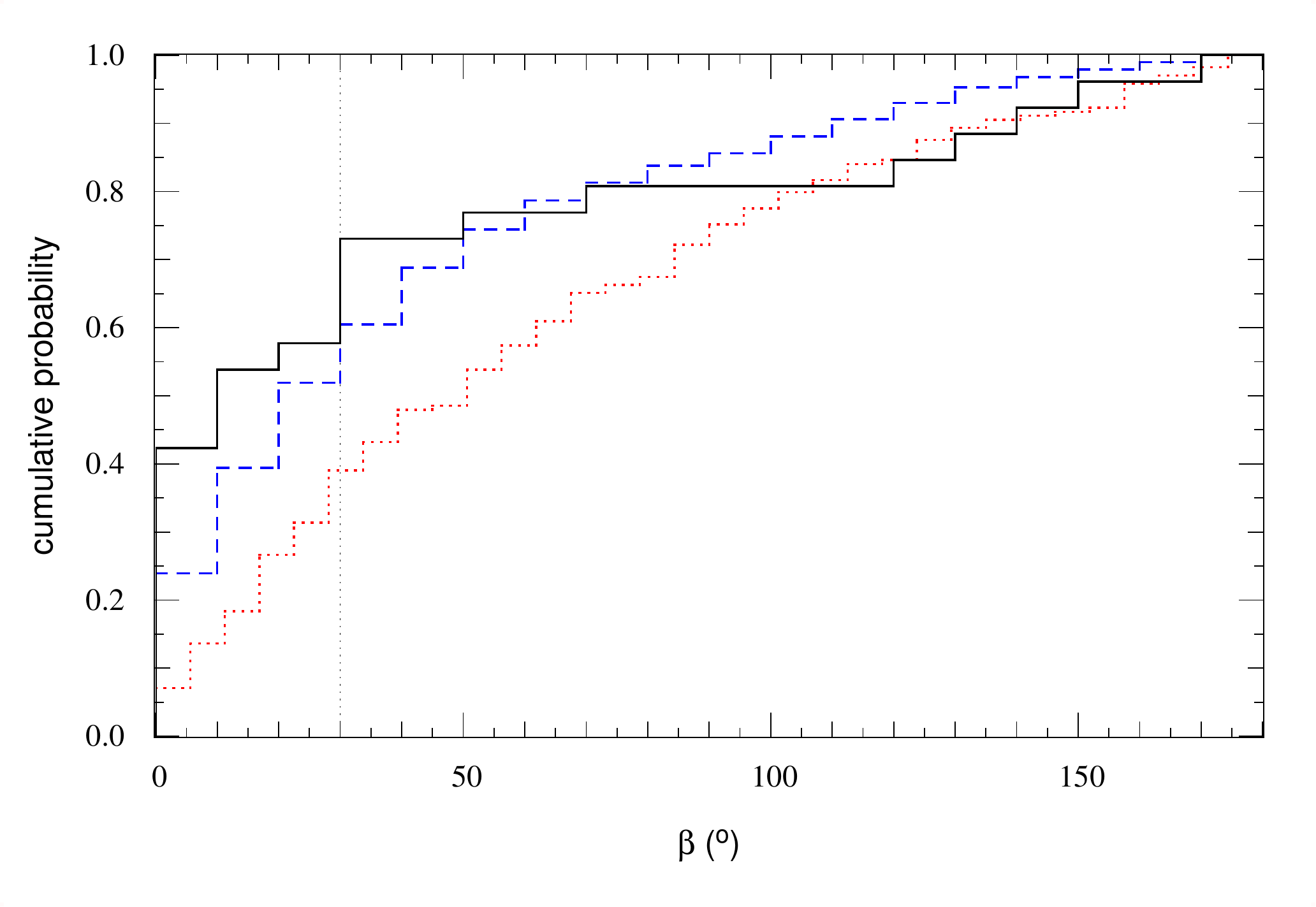}
\caption{\textit{top}: Histogram of all the $\beta$ measured, binned by $20^{\circ}$. \textit{bottom}: Cumulative probability function for models by \citet{Fabrycky:2007p3141} (blued dashed) and \citet{Nagasawa:2008p2997} (red dotted) converted from $\psi$ to $\beta$, compared with current  observations of $\beta$ (plain black). The vertical black dotted line shows $\beta = 30^{\circ}$. Above that, planets are considered misaligned.}\label{fig:hist}

\end{figure}

When combining the 26 RM effects that have been observed, we now see that eight planets are severely misaligned: XO-3b \citep{Hebrard:2008p226,Winn:2009p3777} , HD\,80606b \citep{Moutou:2009p2007, Pont:2009p3828, Winn:2009p3903}, WASP-14b \citep{Johnson:2009p3754}, Hat-P-7b \citep{Winn:2009p3712, Narita:2009p5188}, WASP-8b \citep{Queloz:2010p7085} and WASP-2b, WASP-15b and WASP17b. Of these eight, five have been found to be in retrograde orbits, four from our survey. 

Three additional targets may be misaligned: Kepler-8b \citep{Jenkins:2010p5525}, CoRoT-1b \citep{Pont:2010p6374} and CoRoT-3b \citep{Triaud:2009p3865}. All three are around faint stars and fairly fast rotators making it hard to determine the angle. 
All $\beta$ measurements have been plotted in Fig.~\ref{fig:hist}a. Because we only measure the sky-projection of the angle, the planets can in fact be in a variety of configurations. 
What is their real angle $\psi$ distribution? 

\bigskip

Deconvolving the whole $\beta$ distribution into $\psi$ is hypothesis dependent. 
Hence, to compare the observational data and theoretical predictions we chose to produce cumulative histograms of observational and theoretical $\beta$ angles in Fig.~\ref{fig:hist}b. We transformed predictions from \citet{Fabrycky:2007p3141} and \citet{Nagasawa:2008p2997}, by taking their $\psi$ histograms and transforming them geometrically into observable $\beta$, with the assumption that $I$ is isotropic. For a fixed $\psi$, we define an azimuthal angle $\alpha$ measured from a zero point where the star's north pole is tilted towards the observer. If we precess the star for $\alpha \in [0, 2 \pi[ $ we obtain $\beta$ via a Monte Carlo simulation from solving:

\begin{equation}
\tan \beta \simeq \tan \psi \sin \alpha
\end{equation}
using the conservative assumption that $i = 90^{\circ}$ since these systems are transiting.

Results from this transform are in Fig.~\ref{fig:hist}b. The observational data has been overplotted. Both observations and models by \citet{Fabrycky:2007p3141} agree that about $55\,\%$ of planets should appear with $\beta < 30^{\circ}$\footnote{this criterion of misalignement of $\beta > 30^{\circ}$, is a limit where, with current error bars on $\beta$, one can generally have a significant detection of a misalignement.}. Overall the theoretical distribution is a little steeper than the observations. We clearly remark that predictions by \citet{Nagasawa:2008p2997} agree in range but not in the shape of distribution of observed $\beta$, notably, it lacks enough aligned systems. This model is handy to illustrate the difference between the observations and a distribution isotropic in $\psi$. 

Disc migration models would only produce a steep distribution reaching unity before $30^{\circ}$. A combination of several models is not attempted here because of the vast amount of possibilities and the likelihood that models will evolve.

The theoretical $\psi$ distribution published by \citet{Fabrycky:2007p3141}, transformed into $\beta$, shown along the angle distribution obtained from observations, in Fig.~\ref{fig:hist}b gives a remarkably close match. If the form of this distribution is borne out by future observations, we may then conclude that hot Jupiters are formed by this very mechanism. \citet{Wu:2007p4179} predictions are essentially the same as those from \citet{Fabrycky:2007p3141}.

\bigskip

We also attempted to generalise the method explained in section \ref{sec:psi} to all objects presented in table~\ref{tab:comp}: we are going to assume two distributions for the stellar spin axis in order to derive a distribution of real obliquity $\psi$. Two hypotheses were checked. The first was to assume an isotropy of the stellar axis angle $I$ by taking a uniform distribution in $\cos I$ from 0 to 1; the second was to assume stellar axes are aligned with the plane of the sky. For this last hypothesis 
we assumed all $\cos I $ followed a Gaussian distribution centred on 0 with a variance of 0.1,
an error bar corresponding to the best of what observations can give us at the moment to constrain the stellar $I$.
Taking these hypotheses allows to test for extremes and get an idea of the true proportion of misaligned planetary systems.

\begin{figure}
\centering                     
\includegraphics[width=9cm]{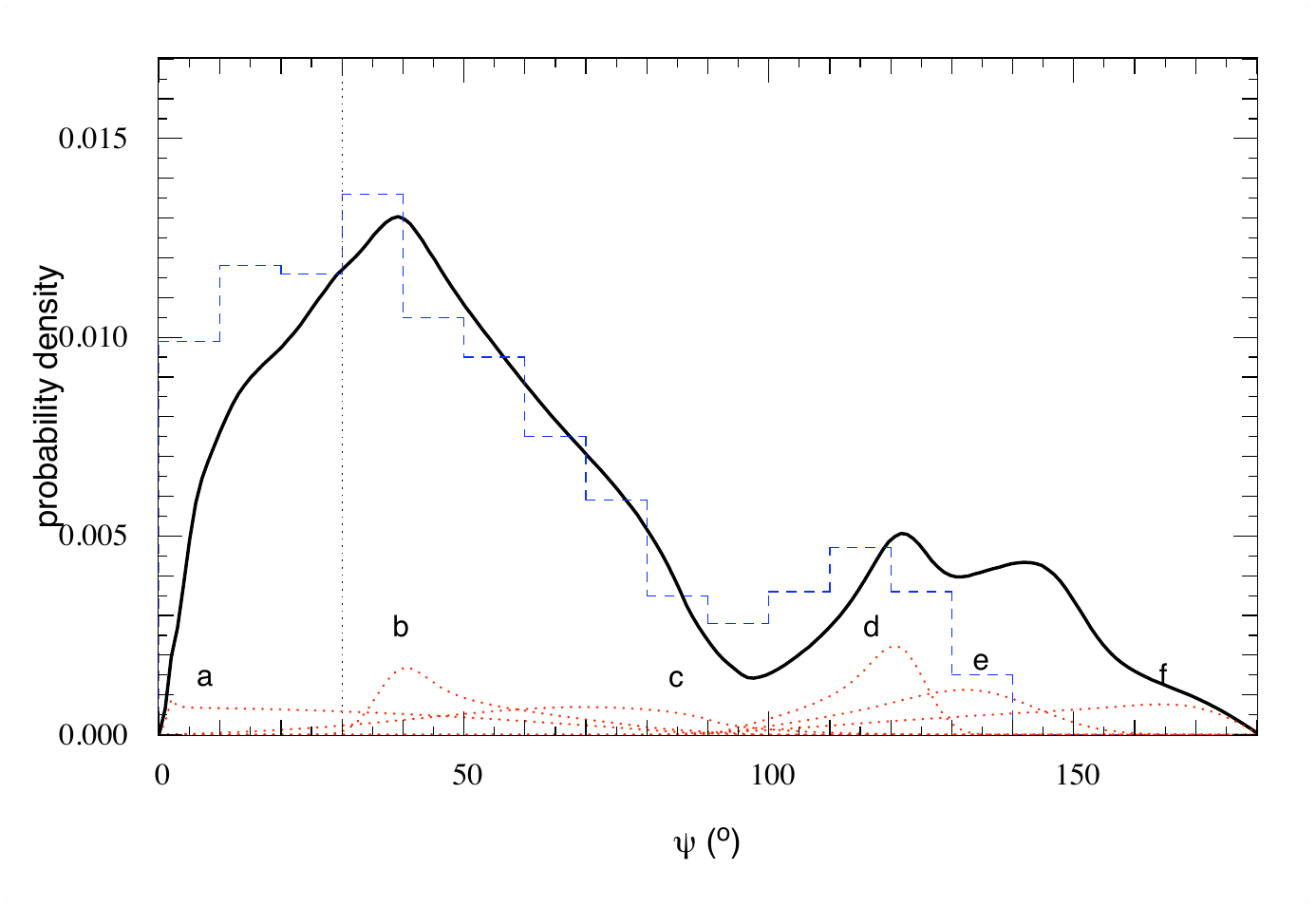}
\caption{The above histogram transformed into the real angle $\psi$ in solid line and smoothed to bins of $1^{\circ}$. Red dotted curves show key individual objects in order to illustrate some of the features of the overall distribution. The blue dashed histogram is the reproduction of the theoretical histogram published by \citet{Fabrycky:2007p3141} and solely plotted over. a)-HD\,189733b  b)-XO-3b  c)-HD\,80606b  d)-WASP-8b   e)-WASP-15b  f)-Hat-P-7b. The black dotted line shows $\psi = 30^{\circ}$. Above that, planets are considered misaligned.}\label{fig:psi}

\end{figure}

The first hypothesis is shown on Fig.~\ref{fig:psi} plotted in comparison with the theoretical predictions by \citet{Fabrycky:2007p3141}. Inferring an isotropic distribution in stellar axes gives us as an upper bound that $86.2\,\%$ of the probability density distribution is at $\psi > 30^{\circ}$. The other hypothesis gives a proportion of $43.6\,\%$ of misaligned systems.
The effect of constraining the stellar $I$ makes every individual contribution narrower in range. Taking a stricter constraint does not change this proportion much.


Both hypotheses are at the extremes of what the real distribution of $I$ is. We interpret these results as showing that between 45 and $85\,\%$ of systems are misaligned with $\psi > 30^{\circ}$. Aligned systems are no longer the norm, radically altering our view on how these hot Jupiters formed. 
\bigskip

\citet{Fabrycky:2007p3141} and \citet{Wu:2007p4179} use the Kozai mechanism \citep{Kozai:1962p3644,Wu:2003p3071} induced by an outer binary companion to the inner planet, to move the planet from the ice line where it is thought to form, to the inner stellar system. As the planet gets closer to the primary, tidal friction helps to break the Kozai cycles and finalise the planet's orbital parameters. Their equations are extracted from work by \citet{Eggleton:2001p6040}. The resulting $\psi$ distribution extends from $0^{\circ}$ up to $150^{\circ}$ away from the primary's rotation axis (see Fig.~\ref{fig:psi}). In this scenario, the planet can be created in a binary star system, or around a single star which acquired a companion through interactions in its cluster of origin \citep{Pfahl:2006p6026}.
\citet{Fabrycky:2007p3141} following on a paper by \citet{Malmberg:2007p3952}, also predict that in multi-planetary systems undergoing Kozai cycles thanks to a nearby star, the most massive planet would survive the resulting planet-planet scattering. 
Although Kozai cycles are usually associated with high eccentricities, we should not be surprised by the presence of so many misaligned planets on circular orbits. As simulated in the case of HD\,80606b in \citet{Fabrycky:2007p3141}, the Kozai cycle has ended (is responsible the close proximity of the planet to the central star at periastron making precession dominated by general relativity rather than by the action of the third body). The planet appears now in a process of circularisation that will take $\sim 0.7$ Gyr, while its angle $\psi$ remains almost constant.

\citet{Nagasawa:2008p2997} model scattering processes between planets creating a pair where one planet is on a close orbit and the other around 40 to 100 AU which then drives Kozai cycles on the inner planet. They also use tidal friction with the star. These authors predict with orbits with a wide range distribution of inclinations and eccentricities which does not reproduce our observations as closely as \citet{Fabrycky:2007p3141} do. All other authors fall short of the wide range of angles that we detect. This, however does not mean that the processes they describe do not happen in combination with those talked about here. In addition we cannot rule out that each of the current theoretical distributions will evolve thanks to greater scrutiny of their starting hypotheses leading to new simulations. Typically, tidal interactions between the star and the planet have been understudied. Any change in the way tidal processes are treated will alter the rate at which planets would realign the stellar spin axis  \citep{Winn:2010p7075}. New effects are also likely to be imagined such as these Kozai cycles between a misaligned planet and a disc presented in \citet{Terquem:2010p6984}.


\bigskip

If the Kozai effect were found to be the dominant process leading to the creation of hot Jupiters, there is no reason why longer period planets should not have undergone similar cycles. The only difference would be that having greater periastron distances, tidal friction was less active. It would then be expected that lone Jupiters on large eccentric orbits be misaligned as well. HD\,80606b would be part of that population. We could then have a \textit{lone Jupiter} population of which \textit{hot Jupiters} are a subset, and another planet population where dynamical interactions and tidal migration did not act.

\begin{sidewaystable*}
\caption{Comparative table of all Rossiter-McLaughlin effects published and parameters. Asterisks show claims of misalignments which remain to be confirmed.}\label{tab:comp}
\begin{tabular}{lllllllll}
\hline
\hline
Planet &$R_\mathrm{p}$ ($R_\mathrm{J}$) & $M_\mathrm{p}$ ($M_\mathrm{J}$) & $ e$ & $P$ (day)& $i$ ($^{\circ}$) & $\beta$ ($^{\circ}$)& $V\sin I$ (km\,s$^{-1}$)&references  \\
\hline
\\
HD\,17156b   & $1.23^{+0.17}_{-0.20}$         & $3.09^{+0.22}_{-0.17}$         & $0.6719^{+0.0052}_{-0.0063}$  & $21.21747^{+0.00070}_{-0.00067}$                    &$85.4^{+1.9}_{-1.2}$        & $-9.4\pm9.3$                   & $6.3\pm1.1$                            & \citet{Cochran:2008p1886, Gillon:2008p767}\\
HD\,80606b   & $0.9\pm0.1$                             & $4.0\pm0.3$                            & $0.934\pm0.003$                          & $111.436\pm0.003$                                                &$89.6\pm0.4$                     & $-53^{+21}_{-34}$          & $2.2$                                        & \citet{Moutou:2009p2007,Winn:2009p3903} \\
HD\,147506b & $0.951^{+0.039}_{-0.053}$  & $8.62^{+0.39}_{-0.55}$         & $0.5163^{+0.0025}_{-0.0023}$ & $5.63341\pm0.00013$                                             &$90.00^{+0.85}_{-0.93}$ & $-0.2^{+12.2}_{-12.5} $  & $22.9^{+1.1}_{-1.2}$              & \citet{Loeillet:2008p881} \\
HD\,149026b & $ 0.718\pm0.065$                  & $0.352\pm0.025$                   & $0$                                                   & $2.87618^{+0.00018}_{-0.00033}$                      &$86.1\pm1.4$                     & $12\pm15$                       & $6.2^{+2.1}_{-0.6}$               & \citet{Wolf:2007p1867} \\
HD\,189733b & $1.178^{+0.016}_{-0.023}$  & $1.138^{+0.022}_{-0.025}$  & $0.0041^{+0.0025}_{-0.0020}$  & $2.21857312^{+0.00000036}_{-0.00000076}$ &$85.51^{+0.10}_{-0.05}$ & $0.85^{+0.28}_{-0.32}$  &  $3.316^{+0.017}_{-0.067}$& \citet{Triaud:2009p3865} \\
HD\,209458b & $1.355 \pm 0.002 $                & $0.657 \pm 0.006$                 & $0.0147 \pm 0.0053$                    & $3.52474859\pm0.00000038$                             &$86.55\pm0.03$                & $4.4\pm1.4$                     & $4.70 \pm 0.16$                    & \citet{Winn:2005p83,Knutson:2007p6005}\\
\\
XO-3b              & $1.5\pm0.2$                            & $12.5\pm1.9$                         &$0.2884\pm0.0035$                       & $3.19161\pm0.00014$                                            &$82.5\pm1.5$                     &$-37.3\pm3.7$                   & $18.31\pm1.3$                      &\citet{Hebrard:2008p226,Winn:2009p3777}  \\
HAT-P-1b       & $1.225\pm0.059$                   & $0.524\pm0.031$                   & $0$                                                   & $4.4652934\pm0.0000093$                                   &$86.28\pm0.20$                &$-3.7\pm2.1$                     & $3.75\pm0.58$                      & \citet{Johnson:2008p1884}\\
HAT-P-7b       & $1.363^{+0.195}_{-0.087}$  &$1.776^{+0.077}_{-0.049}$   & $0$                                                   &$2.2047304\pm 0.0000024$                                   &$80.8^{+2.8}_{-1.2}$        & $-182.5\pm9.4$               &$4.9^{+1.2}_{-0.9}$               & \citet{Pal:2008p4127,Winn:2009p3712,Narita:2009p5188}\\
TrES-1b          & $1.081\pm0.029$                   &$0.61\pm0.06$                         &$0$                                                    &$3.0300722\pm0.0000002$                                    &$88.4\pm0.3$                    &$-30\pm21$                      & $1.3\pm0.3$                           & \citet{Narita:2007p1895}\\
TrES-2b          & $1.220^{+0.045}_{-0.042}$  & $1.198\pm0.053$                   & $0$                                                   & $2.470621\pm0.00017$                                          &$83.62\pm0.14$               &$9\pm12$                          & $1.0\pm0.6$                           & \citet{Winn:2008p1866,Daemgen:2009p5950}\\
TrES-4b	       & $1.674 \pm 0.094$                &$0.84\pm0.10$                         & $0$						&$3.553945\pm 0.000075$				     &$82.81\pm0.33$                &$-7.3 \pm4.6$&$8.3 \pm 1.1$& \citet{Mandushev:2007p5947,Narita:2010p5932} \\
CoRoT-1b*     & $1.49 \pm 0.08$                      &$1.03\pm0.12$                        &$0$                                                    & $1.5089557 \pm 0.0000064$                                 &$85.1\pm0.5$                    &$77\pm 11$                      &$5.2\pm1.0$                            & \citet{Barge:2008p407,Pont:2010p6374} \\
CoRoT-2b      & $1.465\pm0.029$                   & $3.31\pm0.16$                        & $0$                                                   & $1.7429964\pm0.0000017$                                   &$87.84\pm0.10$               &$-7.1\pm5.0$                    &  $11.46^{+0.29}_{-0.44}$    & \citet{Alonso:2008p879, Bouchy:2008p229}\\
CoRoT-3b*     & $0.9934^{+0.0058}_{-0.0058}$  & $21.23^{+0.82}_{-0.59}$ & $0.008^{+0.015}_{-0.005}$        & $4.2567994^{+0.0000039}_{-0.0000031}$         &$86.10^{+0.73}_{-0.52}$& $37.6^{+10.0}_{-22.3}$ & $35.8^{+8.2}_{-8.3}$           & \citet{Triaud:2009p3865}\\
Kepler-8b*      & $1.419^{+0.056}_{-0.058}$ &$0.60^{+0.13}_{-0.19}$           &$0$						 &$3.52254^{+0.00003}_{-0.00005}$                        &$84.07\pm0.33$               &$26.9\pm4.6$   	       & $10.5\pm0.7$                        & \citet{Jenkins:2010p5525}\\
\\
WASP-2b         & $1.077^{+0.055}_{-0.058}$  & $0.866^{+0.076}_{-0.084}$  & $0$                                                   &$2.1522254^{+0.0000015}_{-0.0000014}$           &$84.73^{+0.18}_{-0.19}$& $153^{+11}_{-15}$        & $0.99^{+0.27}_{-0.32}$      & This paper \\
WASP-3b         & $1.29^{+0.05}_{-0.07}$         & $2.04^{+0.07}_{-0.07}$          &$0$                                                   &$1.846834\pm 0.000002$                                         &$84.93^{+1.32}_{-0.80}$& $-3.3^{+4.4}_{-2.5}$       &$14.1^{+1.5}_{-1.3}$          & \citet{Pollacco:2008p5241,Tripathi:2010p6552}\\
WASP-4b         &$1.341^{+ 0.023}_{- 0.029} $& $1.250 ^{+ 0.050}_{- 0.051}$& $0$                                                   & $ 1.3382299 ^{+ 0.0000023}_{- 0.0000021}$    &$89.47^{+0.51}_{-0.24}$  & $-4^{+43}_{-34}$            & $ 2.14 ^{+ 0.38}_{- 0.35}$  & This paper \\
WASP-5b         & $ 1.14^{+ 0.10}_{- 0.04} $     & $1.555 ^{+ 0.066}_{- 0.072}$& $0$                                                   & $ 1.6284229 ^{+ 0.0000043}_{- 0.0000039}$    &$86.1^{+0.7}_{-1.5}$         & $-12.1^{+10.0}_{-8.0}$  & $3.24^{+0.35}_{-0.27}$      & This paper \\
WASP-6b         & $1.224^{+0.051}_{-0.052}$  & $0.503^{+0.019}_{-0.038}$  & $0$                                                   & $3.3610060^{+0.0000022}_{-0.0000035}$         &$88.47^{+0.65}_{-0.47}$  & $11^{+14}_{-18}$           & $1.6^{+0.27}_{-0.17}$         & \citet{Gillon:2009p3869} \\
WASP-8b         & $1.039^{+0.070}_{-0.048}$   & $2.244^{+0.079}_{-0.093}$  & $0.3101^{+0.0029}_{-0.0024}$ & $8.158715^{+0.000016}_{-0.000015}$                &$88.54^{+0.15}_{-0.17}$  & $123.2^{+4.4}_{-3.4}$    & $1.59^{+0.08}_{-0.10}$       & \citet{Queloz:2010p7085} \\
WASP-14b       & $1.281^{+0.075}_{-0.082}$  & $7.34^{+0.51}_{-0.50}$        & $0.0903 \pm 0.0027$                    & $2.23752\pm0.000010$                                          &$84.32^{+0.67}_{-0.57}$  & $33.1\pm7.4$                  & $2.89\pm0.57$                     & \citet{Joshi:2009p2063,Johnson:2009p3754}\\
WASP-15b       & $1.379 ^{+ 0.067}_{-0.058}$& $0.551 ^{+0.041}_{-0.038}$& $0$                                                   & $ 3.752100 ^{+ 0.000009}_{- 0.000011}$             &$85.96^{+0.29}_{-0.41}$ & $ 139.6^{+5.2}_{-4.3}$  & $ 4.27^{+0.26}_{-0.36}$      & This paper \\
WASP-17b      &$1.986 ^{+0.089}_{-0.074}$    &$0.453 ^{+0.043}_{-0.035}$& $0$                                                   &$3.7354330 ^{+0.0000076}_{-0.0000075}$            &$86.63^{+0.39}_{-0.45}$ & $148.5^{+5.1}_{-4.2}$   & $9.92^{+0.40}_{-0.45}$     & This paper \\
WASP-18b       &$1.267 ^{+ 0.062}_{- 0.045}$& $ 10.11 ^{+ 0.24}_{- 0.21}$  & $ 0.00848 ^{+0.00085}_{-0.00095}$ & $0.94145290 ^{+ 0.00000078}_{- 0.00000086}$&$80.6 ^{+1.1}_{-1.3}$        & $ -4.0^{+5.0}_{-5.0}$     & $ 14.61^{+0.81}_{-0.57}$*   & This paper \\
&&\\

\hline

\end{tabular}
\note{* this values is not really a $V\sin I$ but more an amplitude parameter for fitting the Rossiter-McLaughlin effect. Please refer to section \ref{subsec:WASP18} treating WASP-18b}

\end{sidewaystable*}

\section{Conclusions}


The observations reported here bring the total number of transiting planets with known sky-projected obliquities from 20 to 26. Among this enlarged sample, eight show significant projected spin-orbit misalignments; and of these eight, five show apparent retrograde motion. This projected angle $\beta$ can be transformed statistically into the real spin-orbit angle $\psi$. Although $1/3$ of planets have $\beta \neq 0^{\circ}$, the distribution in $\psi$ shows that up to $85\%$ of hot Jupiters are misaligned. The angle range and shape of the overall $\psi$ distribution appears consistent with the predictions of models by \citet{Fabrycky:2007p3141} and \citet{Wu:2007p4179} using the Kozai mechanism to make planets move inwards and tidal friction to reduce their semi-major axis and eventually, circularise them. 

Our evidence therefore points towards a dynamical - not limited to Kozai - and tidal origin for making hot Jupiters so close to their host star. 
This evidence is the strongest yet to suggest that processes others than type I or II migration (using exchange of angular momentum between a planet and a disc) are responsible for the creation of hot Jupiters. Disc migration alone cannot explain the observations; we need to invoke another process. Our interpretation is supported by other facts such as how different hot Jupiters are spread in semi-major axis compared to multiple systems \citep{Wright:2009p6290}, on how lonely hot Jupiters are, and the rarity of hot Jupiters at orbital distances less than two Hill radii from the star \citep{Ford:2006p7047}. 
These results and conclusions should also be a call to account for environmental effects on planetary systems in planet formation simulations. These systems are not in isolation and interact with their neighbours.


We are seeing the coming of a new diversity in planetary parameters, coming after large diversities in mass, period, eccentricity and radius. The variety of angles $\beta$, transformed into $\psi$, is an indication of the physical processes that happened before, during and after planet formation. Once again the measurement of a new observable has brought a large variety of values reflecting how rich nature is. 

As more transiting systems are discovered in wide-field surveys, and follow-up observations of the kind reported here are made, the statistical picture that is beginning to emerge will become clearer.

\begin{acknowledgements} 

The authors would like to acknowledge the use of ADS and \textit{Simbad} but foremost, the help and the kind attention of the ESO staff at La Silla.

As a small show of our immense gratitude, we would like to underline the enormous work and constant support, upkeep and maintenance of the \textit{Euler} observing station at La Silla, as well as the quick and useful help given by engineers, mechanical and IT staff from the Observatoire de Gen\`eve. Without you, a lot of this work would not have been possible. This station is like a jewel, thanks!   

An extensive use of the \texttt{exoplanet.eu} encyclopaedia was made; many thanks to Jean Schneider for the continuous update and upkeep of this database. 

We would also like to thank G. Ogilvie, M. Davies, A. Barker, D. Malmberg, R. Alonso, S. Matsumura, G. Laughlin, D. Fabrycky, T. Guillot and E. Ford for their helpful and interesting conversations and to the battery of observers that helped obtain the data on HARPS and CORALIE, notably X. Bonfils, G. Chauvin, G. H\'ebrard, G. Lo Curto, C. Lovis, M. Marmier, C. Moutou and D. Naef. 

Our final thanks go to Josh Winn, our referee, who helped greatly to refine this paper, its results and their interpretation. 

This work is supported by the Swiss Fond National de Recherche Scientifique.
  
\end{acknowledgements} 

\bibliographystyle{aa}
\bibliography{bibtex}

\appendix
\section{Comparative tables for each star}\label{sec:comp}

Here, for transparency, are the tables recording the results from the various fits that were done for each star, which, par comparing them, led to the choice of our solutions. $\chi^2$ have been tabulated only for the radial velocity data that was used for our analysis. In addition, to show where the most important contributions come from, $\chi^2$ have also been added for each set of radial velocities separately, as they are presented in the Journal of Observations in the following appendix. Finally, because our aim was to measure $\beta$, a line with the $\chi^2$ only during the Rossiter-McLaughlin effect has been added. Comparisons between the $\chi^2$ contributions of the overall reflex motion of the star with contributions during the Rossiter-McLaughlin will show that we tend to fit better during transit than outside. The number of points during the Rossiter-McLaughlin effect have been chosen as all the points measured during transit, plus one point immediately on either side when available.


\begin{table*}
\caption{Differences between fits of WASP-2b, 4b \& 5b. $\chi^2_\mathrm{reduced}$ has been estimated for the radial velocities only.}\label{tab:WASP1comp}
\begin{tabular}{lllllll}
WASP-2b\\
\\
\hline
\hline
&&&&& Fixed Photometry\\
$V\sin I$ Prior &  on & off & on & off &off&\\
\hline
\\
$V\sin I$ (km\,s$^{-1}$)  &$ 1.08 ^{+ 0.26}_{- 0.31}$ &$0.99^{+0.27}_{-0.32}$     &$1.02^{+0.28}_{-0.25}$               & $0.93^{+0.26}_{-0.30}$& $0.99^{+0.29}_{-0.33}$\\
$\beta$ ($^{\circ}$)           &$154^{+10}_{-12}$            &$153^{+11}_{-15}$   &$145^{+12}_{-15}$                       & $143^{+12}_{-18}$& $152^{+12}_{-16}$\\
$e$                                      &-                                          &-                                 &$0.035^{+0.016}_{-0.014}$ &$0.036^{+0.017}_{-0.015}$&-\\
$\omega$ ($^{\circ}$)      &-                                          &-                                 &$-103^{+6}_{-12}$                      &$-103^{+6}_{-11}$&-\\
\\
\hline
\\
\multicolumn{5}{l}{\textit{all 58 RVs, 3 sets}}\\
$\chi^2_\mathrm{RV}$ &$100.6\pm14.2$         &$100.5\pm14.2$          &$93.2\pm13.6$             &$92.9\pm13.6$&$100.5\pm14.2$\\
$N_\mathrm{dof}$         &47                                  &47                                   & 45                                   & 45                     &47\\
$\chi^2_\mathrm{reduced}$ &$2.14\pm0.30$         &$2.14\pm0.30$          &$2.07\pm0.30$             &$2.06\pm0.30$&$2.14\pm0.30$\\
\\
$\chi^2_\mathrm{SOPHIE,\,8\,RVs}$      &$28.1\pm7.5$         &$27.9\pm7.5$          &$26.9\pm7.3$             &$27.0\pm7.4$  &$27.9\pm7.5$\\
$\chi^2_\mathrm{CORALIE,\,20\,RVs}$ &$15.6\pm5.6$         &$15.5\pm5.6$          &$21.8\pm6.6$             &$21.4\pm6.5$  &$15.7\pm5.6$\\
$\chi^2_\mathrm{HARPS,\,30\,RVs}$     &$56.9\pm10.7$      &$57.1\pm10.7$        &$44.4\pm9.4$             &$44.5\pm9.4$   &$57.0\pm10.7$\\
\\
$\chi^2_\mathrm{HARPS,\,RM, \,17\,RVs}$&$20.7\pm6.4$      &$20.7\pm6.4$        &$20.9\pm6.5$            &$20.7\pm6.4$    &$20.7\pm6.4$\\
\\
\hline
\\
\\
\hline
\\
 &  no RM & RM fixed & RM fixed & no RM &  RM fixed & RM fixed\\
\hline
\\
$V\sin I$ (km\,s$^{-1}$)  &-           &1.6  &0.9   &-              & 1.6    & 0.9\\
$\beta$ ($^{\circ}$)           &-            &0      &0       &-                       & 0 & 0\\
$e$                                      &-                                          &-           &-                      &$0.041^{+0.015}_{-0.016}$ &$0.044^{+0.016}_{-0.014}$&$0.044^{+0.014}_{-0.016}$\\
$\omega$ ($^{\circ}$)      &-                                          &-            &-                     &$-96^{+5}_{-6}$                      &$-98^{+5}_{-6}$&$-97^{+5}_{-6}$\\
\\
\hline
\\
\multicolumn{5}{l}{\textit{all 58 RVs, 3 sets}}\\
$\chi^2_\mathrm{RV}$           &$113.7\pm15.1$         &$164.0\pm18.1$   &  $135.8\pm16.5$     &$105.7\pm14.5$             &$154.0\pm17.6$ &$126.4\pm15.9$\\
$N_\mathrm{dof}$                  &49                                  &47                             & 47                               & 47                                     &45                          &45 \\
$\chi^2_\mathrm{reduced}$ &$2.32\pm0.31$         &$3.49\pm0.39$         &  $2.89\pm0.35$       &$2.25\pm0.31$               &$3.42\pm0.39$   &$ 2.81\pm0.35$\\
\\
$\chi^2_\mathrm{SOPHIE,\,8\,RVs}$      &$31.0\pm7.9$         &$29.3\pm7.7$          &$30.1\pm7.7$             &$31.3\pm7.9$      &$30.4\pm7.8$&   $30.8\pm7.8$\\
$\chi^2_\mathrm{CORALIE,\,20\,RVs}$ &$15.7\pm5.6$         &$15.4\pm5.5$          &$15.4\pm5.6$             &$21.5\pm6.6$      &$23.0\pm6.8$&   $22.8\pm6.8$\\
$\chi^2_\mathrm{HARPS,\,30\,RVs}$     &$67.0\pm11.6$      &$119.4\pm15.5$      &$90.4\pm13.4$          &$52.8\pm10.3$   &$100.6\pm14.2$&$72.8\pm12.1$\\
\\
$\chi^2_\mathrm{HARPS,\,RM, \,17\,RVs}$&$30.8\pm7.8$      &$82.7\pm12.9$      &$53.9\pm10.4$         &$29.0\pm7.6$    &$79.1\pm12.6$    &$51.1\pm10.1$\\
\\
\hline
\\
\\
WASP-4b\\
\\
\hline
\hline
 &&&&& Fixed Photometry\\
$V\sin I$ Prior &  on & off & on & off &on\\
\hline
\\
$V\sin I$ (km\,s$^{-1}$)  &$ 2.14 ^{+ 0.38}_{- 0.35}$ &$4^{+46}_{-2}$     &$2.15^{+0.45}_{-0.39}$               & $78^{+41}_{-75}$ &$2.19^{+0.35}_{-0.45}$\\
$\beta$ ($^{\circ}$)           &$-4^{+43}_{-34}$            &$4^{+84}_{-80}$   &$0.^{+34}_{-41}$                       & $28^{+118}_{-0}$&$-5^{+39}_{-38}$\\
$e$                                      &-                                          &-                                 &$0.0105^{+0.0036}_{-0.0072}$ &$0.0106^{+0.0038}_{-0.0074}$&-\\
$\omega$ ($^{\circ}$)      &-                                          &-                                 &$-108^{+282}_{-58}$                      &$-107^{+280}_{-61}$&-\\
\\
\hline
\\
\multicolumn{5}{l}{\textit{all 56 RVs, 2 sets}}\\
$\chi^2_\mathrm{RV}$           &$77.8\pm12.5$         &$78.0\pm12.5$          &$75.3\pm12.4$             &$75.3\pm12.3$&$77.8\pm12.5$\\
$N_\mathrm{dof}$         &46                                  &46                                   & 44                                   & 44                     &46\\
$\chi^2_\mathrm{reduced}$ &$1.69\pm0.27$         &$1.70\pm0.27$          &$1.71\pm0.28$             &$1.71\pm0.28$&$1.69\pm0.27$\\
\\
$\chi^2_\mathrm{CORALIE,\,24\,RVs}$ &$28.4\pm7.5$         &$29.1\pm7.6$          &$27.6\pm7.4$             &$28.0\pm7.5$  &$28.6\pm7.6$\\
$\chi^2_\mathrm{HARPS,\,32\,RVs}$     &$49.4\pm9.9$         &$48.9\pm9.9$          &$47.8\pm9.8$             &$47.3\pm9.7$   &$49.2\pm9.9$\\
\\
$\chi^2_\mathrm{HARPS,\,RM, \,15\,RVs}$&$12.6\pm5.0$      &$12.5\pm5.0$        &$12.2\pm4.9$            &$12.0\pm4.9$    &$12.7\pm5.0$\\
\\
\hline

\end{tabular}
\end{table*}

\begin{table*}
\caption{Differences between fits of WASP-5b \& 15b. $\chi^2_\mathrm{reduced}$ has been estimated for the radial velocities only.}\label{tab:WASP2comp}
\begin{tabular}{llllll}
\\
WASP-5b\\
\\
\hline
\hline
 &&&&& Fixed Photometry\\
$V\sin I$ Prior &  on & off & on & off &off\\
\hline
\\
$V\sin I$ (km\,s$^{-1}$)  &$3.24^{+0.35}_{-0.27}$ &$3.24^{+0.34}_{-0.35}$     &$3.32^{+0.30}_{-0.32}$               & $3.36^{+0.32}_{-0.46}$ &$3.18^{+0.26}_{-0.31}$\\
$\beta$ ($^{\circ}$)           &$-12.1^{+10.0}_{-8.0}$            &$-12.4^{+11.9}_{-8.2}$   &$-14.1^{+10.8}_{-7.8}$                       & $-16.1^{+14.2}_{-9.3}$ &$-12.0^{+7.7}_{-7.3}$\\
$e$                                      &-                                          &-                                 &$0.0209^{+0.0081}_{-0.0075}$ &$0.0209^{+0.0071}_{-0.0087}$&-\\
$\omega$ ($^{\circ}$)      &-                                          &-                                 &$-137^{+14}_{-16}$                      &$-137^{+12}_{-17}$&-\\
\\
\hline
\\
\multicolumn{5}{l}{\textit{all 49 RVs, 2 sets}}\\
$\chi^2_\mathrm{RV}$           &$143.7\pm17.0$         &$144.3\pm17.0$          &$136.8\pm16.5$             &$136.7\pm16.5$ & $145.1\pm17.0$\\
$N_\mathrm{dof}$         &39                                  &39                                   & 37                                   & 37                     &39\\
$\chi^2_\mathrm{reduced}$ &$3.69\pm0.43$         &$3.70\pm0.44$          &$3.70\pm0.45$             &$3.70\pm0.45$ & $3.72\pm0.44$\\
\\
$\chi^2_\mathrm{CORALIE,\,16\,RVs}$ &$20.4\pm6.4$         &$20.5\pm6.4$          &$26.3\pm7.2$             &$26.0\pm7.2$       &$20.4\pm6.4$\\
$\chi^2_\mathrm{HARPS,\,33\,RVs}$     &$123.3\pm15.7$    &$123.8\pm15.7$     &$110.6\pm14.9$        &$110.7\pm14.9$   &$124.8\pm15.8$\\
\\
$\chi^2_\mathrm{HARPS,\,RM, \,15\,RVs}$&$42.8\pm9.2$   &$43.9\pm9.4$        &$40.6\pm9.0$              &$40.7\pm9.0$    &$43.9\pm9.4$\\
\\
\hline
\\
\\
WASP-15b\\
\\
\hline
\hline
 &&&&& Fixed Photometry\\
$V\sin I$ Prior &  on & off & on & off &off\\
\hline
\\
$V\sin I$ (km\,s$^{-1}$)  &$4.26^{+0.27}_{-0.32}$ &$4.27^{+0.26}_{-0.36}$     &$4.37^{+0.29}_{-0.32}$               & $4.36^{+0.27}_{-0.34}$       &$4.26^{+0.28}_{-0.31}$\\
$\beta$ ($^{\circ}$)           &$139.8^{+5.1}_{-4.5}$   &$139.6^{+5.2}_{-4.3}$      &$142.6^{+5.3}_{-4.5}$                  & $142.7^{+5.3}_{-5.0}$         &$139.7^{+4.0}_{-4.0}$\\
$e$                                      &-                                          &-                                            &$0.043^{+0.020}_{-0.022}$         &$0.043^{+0.022}_{-0.023}$&-\\
$\omega$ ($^{\circ}$)      &-                                          &-                                             &$96^{+45}_{-22}$                         &$96^{+38}_{-26}$                 &-\\
\\
\hline
\\
\multicolumn{5}{l}{\textit{all 95 RVs, 2 sets}}\\
$\chi^2_\mathrm{RV}$          &$133.1\pm16.3$         &$133.3\pm16.3$          &$130.3\pm16.1$             &$130.1\pm16.1$&$133.1\pm16.3$\\
$N_\mathrm{dof}$                  &85                                  &85                                   & 83                                   & 83                          &85\\
$\chi^2_\mathrm{reduced}$ &$1.57\pm0.19$         &$1.57\pm0.19$               &$1.57\pm0.19$             &$1.57\pm0.19$   &$1.57\pm0.19$\\
\\
$\chi^2_\mathrm{CORALIE,\,44\,RVs}$ &$53.7\pm10.4$         &$53.4\pm10.3$    &$53.5\pm10.3$             &$54.4\pm10.4$       &$53.9\pm10.4$\\
$\chi^2_\mathrm{HARPS,\,51\,RVs}$     &$79.4\pm12.6$        &$79.8\pm12.6$     &$76.7\pm12.4$        &$75.8\pm12.3$   &$79.2\pm12.6$\\
\\
$\chi^2_\mathrm{HARPS,\,RM, \,33\,RVs}$&$47.3\pm9.7$   &$47.3\pm9.7$        &$46.9\pm9.7$              &$46.5\pm9.6$    &$47.1\pm9.7$\\
\\
\hline
\\
\end{tabular}
\end{table*}

\begin{table*}
\caption{Differences between fits of WASP-17b \& 18b. $\chi^2_\mathrm{reduced}$ has been estimated for the radial velocities only.}\label{tab:WASP3comp}
\begin{tabular}{llllll}
\\
WASP-17b\\
\\
\hline
\hline
 &&&&& Fixed Photometry\\
$V\sin I$ Prior &  on & off & on & off &off\\
\hline
\\
$V\sin I$ (km\,s$^{-1}$)  &$9.92^{+0.40}_{-0.45}$ &$10.14^{0.58}_{-0.79}$     &$9.95^{+0.45}_{-0.43}$               & $10.27^{+0.68}_{-0.84}$         & $10.03^{+0.63}_{-0.63}$\\
$\beta$ ($^{\circ}$)           &$148.5^{+5.1}_{-4.2}$   &$147.3^{+5.9}_{-5.5}$      &$150.9^{+5.2}_{-5.9}$                  & $150.5^{+6.1}_{5.7}$               & $147.5^{+4.2}_{-4.0}$\\
$e$                                      &-                                          &-                                            &$0.062^{+0.024}_{-0.039}$        &$0.066^{+0.030}_{-0.043}$     &-\\
$\omega$ ($^{\circ}$)      &-                                          &-                                            &$34^{+34}_{-72}$                         &$45^{+30}_{-77}$                       &-\\
\\
\hline
\\
\multicolumn{5}{l}{\textit{all 124 RVs, 4 sets}}\\
$\chi^2_\mathrm{RV}$           &$190.1\pm19.5$         &$190.4\pm19.5$          &$187.3\pm19.4$             &$186.9\pm19.3$     &$191.6\pm19.6$\\
$N_\mathrm{dof}$                  &112                                &112                                 & 110                                   & 110                          &112\\
$\chi^2_\mathrm{reduced}$ &$1.70\pm0.17$            &$1.70\pm0.17$            &$1.70\pm0.17$                &$1.70\pm0.17$      &$1.71\pm0.17$\\
\\
$\chi^2_\mathrm{CORALIE,\,49\,RVs}$ &$47.6\pm9.8$         &$47.4\pm9.7$      &$47.2\pm9.7$          &$47.7\pm9.8$       &$47.5\pm9.7$\\
$\chi^2_\mathrm{CORALIE,\,15\,RVs}$ &$15.0\pm5.5$         &$15.0\pm5.5$     &$16.2\pm5.7$           &$16.9\pm5.8$       &$15.0\pm5.5$\\
$\chi^2_\mathrm{HARPS,\,16\,RVs}$     &$23.6\pm6.9$        &$23.7\pm6.9$      &$23.5\pm6.9$           &$23.7\pm6.9$   &$23.9\pm6.9$\\
$\chi^2_\mathrm{HARPS,\,44\,RVs}$     &$103.8\pm14.4$   &$104.3\pm14.4$  &$100.4\pm14.2$      &$98.6\pm14.0$   &$105.2\pm14.5$\\
\\
$\chi^2_\mathrm{CORALIE,\,RM, \,13\,RVs}$&$9.8\pm4.4$   &$9.5\pm4.4$        &$9.9\pm4.4$              &$10.1\pm4.5$    &$9.8\pm4.4$\\
$\chi^2_\mathrm{HARPS,\,RM, \,28\,RVs}$&$59.3\pm10.9$  &$59.8\pm10.9$   &$59.4\pm10.9$          &$58.7\pm10.8$    &$60.7\pm11.0$\\
\\
\hline
\\
\\
WASP-18b\\
\\
\hline
\hline
 &&&&& Fixed Photometry\\
$V\sin I$ Prior &  on & off & on & off & off\\
\hline
\\
$V\sin I$ (km\,s$^{-1}$) * &$14.04^{+0.73}_{-0.52}$ *&$14.66^{0.86}_{-0.58} *$   &$14.67^{+0.81}_{-0.57}$           *  & $15.57^{+1.01}_{-0.69}$ *          & $15.59^{+0.56}_{-0.57}$ *\\
$\beta$ ($^{\circ}$)           &$-11.1^{+6.6}_{-5.8}$        &$-10.1^{+6.2}_{-5.8}$         &$-5.0^{+6.2}_{-5.6}$                       & $-4.0^{+5.0}_{-5.0}$                     & $-4.2^{+4.6}_{-4.6}$\\
$e$                                      &-                                             &-                                               &$0.0084^{+0.0008}_{-0.0010}$   &$0.0085^{+0.0009}_{-0.00010}$ &$0.0085^{+0.0010}_{-0.0010}$\\
$\omega$ ($^{\circ}$)      &-                                              &-                                              &$-92.8^{+5.2}_{-3.9}$                     &$-92.1^{+4.9}_{-4.3}$                    &$-92.5^{+2.7}_{-3.0}$\\
\\
\hline
\\
\multicolumn{5}{l}{\textit{all 60 RVs, 2 sets}}\\
$\chi^2_\mathrm{RV}$           &$283.3\pm23.8$         &$279.3\pm23.6$          &$179.7\pm18.9$             &$177.8\pm18.9$&$178.4\pm18.9$\\
$N_\mathrm{dof}$                  &50                                  &50                                   & 48                                     & 48                         &48\\
$\chi^2_\mathrm{reduced}$ &$5.67\pm0.48$            &$5.58\pm0.47$            &$3.74\pm0.39$                &$3.70\pm0.39$  &$3.72\pm0.39$\\
\\
$\chi^2_\mathrm{CORALIE,\,37\,RVs}$ &$132.4\pm16.3$         &$131.2\pm16.2$    &$69.2\pm11.8$             &$66.7\pm11.5$       &$67.5\pm10.4$\\
$\chi^2_\mathrm{HARPS,\,23\,RVs}$     &$150.9\pm17.4$        &$148.1\pm17.2$     &$110.5\pm14.9$        &$111.1\pm14.9$   &$110.9\pm14.9$\\
\\
$\chi^2_\mathrm{HARPS,\,RM, \,12\,RVs}$&$113.7\pm15.1$   &$110.2\pm14.8$        &$98.7\pm14.0$              &$98.0\pm14.0$    &$98.6\pm14.0$\\
\\
\hline

\end{tabular}
\note{* these values are not really $V\sin I$ but more an amplitude parameter for fitting the Rossiter-McLaughlin effect. Please refer to section \ref{subsec:WASP18} treating WASP-18b}
\end{table*}

\section{Journal of Observations}

The Radial-Velocity data extracted by fitting a Gaussian function on a Cross-Correlation Function resulting from comparing the spectra with a mask corresponding to its spectral type. The data is presented per instrument and separated in various datasets: overall Doppler shift and Rossiter-McLaughlin effect, as was done for the fits. Within each dataset, it is presented chronologically. The data is available for download at CDS - Strasbourg.

\begin{table}
\caption{RV data for WASP-2b.}\label{tab:WASP2data}
\begin{tabular}{llll}
\hline
\hline
bjd (-\,2\,450\,000)&RV (km\,s$^{-1}$)& $\sigma_\mathrm{RV}$ (km\,s$^{-1}$) & $t_\mathrm{exp}$ (s) \\
\hline
\\
SOPHIE&\multicolumn{2}{l}{\textit{orbital Doppler shift}}\\
\\
3982.3786&-27.711&0.012 & 2500\\
3982.4962&-27.736&0.013 & 2500\\
3991.3817&-27.780&0.011 & 1200\\
3991.5102&-27.812&0.011 & 1200\\
3996.3529&-28.037&0.011 & 1200\\
3996.4301&-28.020&0.012 & 1200\\
3997.3824&-27.723&0.012 & 1500\\
3998.3415&-27.987&0.011 & 1200\\
\\
CORALIE &\multicolumn{2}{l}{\textit{orbital Doppler shift}}\\
\\
4764.504689&-27.85981&0.01932 & 1801\\
4765.509206&-27.65742&0.02299 & 1801\\
4766.508375&-27.84650&0.01380 & 1801\\
4769.512698&-27.78332&0.01319 & 1801\\
4770.508244&-27.70918&0.01711 & 1801\\
4771.509646&-27.82787&0.01391 & 1801\\
4772.519878&-27.64359&0.01484 & 1801\\
4773.513595&-27.88050&0.01411 & 1801\\
4774.515784&-27.61010&0.01413 & 1801\\
4775.512417&-27.90960&0.01485 & 1801\\
4776.512326&-27.61153&0.02040 & 1801\\
5001.827721&-27.81832&0.01152 & 1801\\
5013.750160&-27.71932&0.01163 & 1801\\
5037.699667&-27.85913&0.01200 & 1801\\
5038.721568&-27.68417&0.00961 & 1801\\
5039.689751&-27.79590&0.01101 & 1801\\
5041.718024&-27.72765&0.01018 & 1801\\
5042.650566&-27.84181&0.00947 & 1801\\
5092.543315&-27.70924&0.01325 & 1801\\
5097.558717&-27.68199&0.01203 & 1801\\
\\
HARPS&\multicolumn{2}{l}{\textit{Rossiter-McLaughlin effect}}\\
\\
4754.528495&-27.70390&0.00702 & 900\\
4754.620408&-27.66549&0.00477& 900\\
4755.490831&-27.71635&0.00649 & 400\\
4755.495101&-27.72204&0.00675 & 400\\
4755.500251&-27.72739&0.00458 & 300\\
4755.505239&-27.72374&0.00441 & 300\\
4755.509984&-27.72192&0.00491 & 400\\
4755.515331&-27.72547&0.00650 & 400\\
4755.520226&-27.72620&0.00469 & 400\\
4755.525087&-27.72943&0.00461 & 400\\
4755.530133&-27.74799&0.00516 & 400\\
4755.535179&-27.73390&0.00495 & 400\\
4755.540074&-27.73202&0.00498 & 400\\
4755.545178&-27.73969&0.00502 & 400\\
4755.550166&-27.72911&0.00498 & 400\\
4755.555073&-27.72841&0.00500 & 400\\
4755.560015&-27.74669&0.00595 & 400\\
4755.565061&-27.74516&0.00671 & 400\\
4755.570164&-27.73728&0.00604 & 400\\
4755.575164&-27.74225&0.00656 & 400\\
4755.580025&-27.74699&0.00679 & 400\\
4755.585070&-27.75745&0.00643 & 400\\
4755.590070&-27.74940&0.00709 & 400\\
4755.595197&-27.76458&0.00730 & 400\\
4755.602187&-27.77178&0.00522 & 600\\
4755.609420&-27.75318&0.00492 & 600\\
4755.616885&-27.75791&0.00581 & 600\\
4755.624130&-27.77039&0.00506 & 600\\
4756.503729&-27.79724&0.00242 & 900\\
4756.556086&-27.77834&0.00285 & 900\\
\\
\hline
\end{tabular}
\end{table}

\begin{table}
\caption{RV data for WASP-4b.}\label{tab:WASP4data}
\begin{tabular}{lllll}
\hline
\hline
bjd (-\,2\,450\,000)&RV (km\,s$^{-1}$)& $\sigma_\mathrm{RV}$ (km\,s$^{-1}$)& $t_\mathrm{exp}$ (s)\\
\hline

\\
CORALIE &\multicolumn{2}{l}{\textit{orbital Doppler shift}}\\
\\

4359.710824&57.57468&0.02053 & 1801\\
4362.631217&57.80604&0.02094 & 1801\\
4364.652602&57.68875&0.02460 & 1801\\
4365.736906&57.95170&0.01778 & 1801\\
4372.757994&57.59277&0.01652 & 1801\\
4376.688827&57.64569&0.01605 & 1801\\
4378.668871&57.79988&0.01458 & 1801\\
4379.736306&57.52086&0.01624 & 1801\\
4380.610348&57.78459&0.01437 & 1801\\
4382.790257&57.87592&0.02015 & 1801\\
4383.552773&57.50732&0.01570 & 1801\\
4387.619048&57.51081&0.01611 & 1801\\
4408.661101&57.79088&0.01875 & 1801\\
4409.519323&57.84236&0.02407 & 1801\\
4720.589143&57.73974&0.02532 & 1801\\
4722.674920&57.90628&0.01727 & 1801\\
4725.569004&58.00982&0.02329 & 1801\\
4729.588760&57.98945&0.01967 & 1801\\
4730.646040&57.82448&0.01753 & 1801\\
4760.593166&57.84475&0.01749 & 1801\\
4761.639625&57.99209&0.01691 & 1801\\
4762.654627&57.70943&0.01785 & 1801\\
4763.603116&57.52516&0.01753 & 1801\\
4763.725804&57.48750&0.02455 & 1801\\
\\
HARPS&\multicolumn{2}{l}{\textit{Rossiter-McLaughlin effect}}\\
\\
4747.809113&57.61982&0.00597 & 900\\
4748.501785&57.94526&0.00544 & 900\\
4748.552731&57.89782&0.00623 & 600\\
4748.560173&57.88884&0.00616 & 600\\
4748.567684&57.88212&0.00639 & 600\\
4748.575207&57.87224&0.00661 & 600\\
4748.582857&57.87643&0.00716 & 600\\
4748.590241&57.83952&0.00705 & 600\\
4748.597752&57.84293&0.00710 & 600\\
4748.605332&57.83795&0.00672 & 600\\
4748.612716&57.86116&0.00730 & 600\\
4748.620308&57.85916&0.00703 & 600\\
4748.627958&57.84646&0.00747 & 600\\
4748.635273&57.81970&0.00725 & 600\\
4748.642865&57.81142&0.00671 & 600\\
4748.650446&57.79059&0.00649 & 600\\
4748.657957&57.76719&0.00594 & 600\\
4748.665329&57.75065&0.00574 & 600\\
4748.672910&57.73640&0.00618 & 600\\
4748.680421&57.73107&0.00673 & 600\\
4748.688001&57.71750&0.00640 & 600\\
4748.695374&57.74694&0.00602 & 600\\
4748.703024&57.72254&0.00621 & 600\\
4748.710546&57.71510&0.00564 & 600\\
4748.718069&57.71593&0.00573 & 600\\
4748.725511&57.69356&0.00562 & 600\\
4748.733034&57.68893&0.00570 & 600\\
4748.740545&57.70283&0.00599 & 600\\
4748.748056&57.69677&0.00603 & 600\\
4748.755579&57.67217&0.00624 & 600\\
4748.763090&57.68511&0.00652 & 600\\
4750.744196&57.89058&0.00405 & 1800\\

\\
\hline
\end{tabular}
\end{table}

\begin{table}
\caption{RV data for WASP-5b.}\label{tab:WASP5data}
\begin{tabular}{lllll}
\hline
\hline
bjd (-\,2\,450\,000)&RV (km\,s$^{-1}$)& $\sigma_\mathrm{RV}$ (km\,s$^{-1}$)& $t_\mathrm{exp}$ (s)\\
\hline

\\
CORALIE &\multicolumn{2}{l}{\textit{orbital Doppler shift}}\\
\\

4359.614570&19.78859&0.01785 & 1801\\
4362.654785&19.94934&0.01630 & 1801\\
4364.676219&19.76409&0.02288 & 1801\\
4365.682733&20.20817&0.01378 & 1801\\
4372.781658&19.75922&0.01459 & 1801\\
4374.817984&20.01631&0.04516 & 1801\\
4376.714797&20.26098&0.01558 & 1801\\
4377.762440&19.75519&0.01445 & 1801\\
4379.627523&19.91862&0.01587 & 1801\\
4380.682557&19.84080&0.01142 & 1801\\
4387.644986&19.82288&0.01649 & 1801\\
4720.835375&20.05679&0.01816 & 1801\\
4724.603499&19.73117&0.01661 & 1801\\
4732.745208&19.74646&0.01812 & 1802\\
4733.686048&20.16201&0.01556 & 1801\\
4734.694930&19.99692&0.01726 & 1801\\
\\
HARPS&\multicolumn{2}{l}{\textit{Rossiter-McLaughlin effect}}\\
\\
4749.722428&20.29554&0.00334 & 1200\\
4750.766740&19.83105&0.00361 & 1800\\
4753.696462&19.79468&0.00391 & 1800\\
4754.773661&20.25200&0.00338 & 1800\\
4756.491182&20.18379&0.00363 & 900\\
4756.570483&20.11309&0.00456 & 600\\
4756.578469&20.11813&0.00483 & 600\\
4756.586466&20.09061&0.00527 & 600\\
4756.594846&20.08995&0.00551 & 600\\
4756.602762&20.08266&0.00497 & 600\\
4756.610747&20.07852&0.00478 & 600\\
4756.618814&20.06709&0.00485 & 600\\
4756.626904&20.05854&0.00512 & 600\\
4756.635121&20.06323&0.00462 & 600\\
4756.643107&20.06533&0.00456 & 600\\
4756.651173&20.06765&0.00439 & 600\\
4756.659090&20.05028&0.00496 & 600\\
4756.667399&20.02693&0.00544 & 600\\
4756.675223&20.01679&0.00543 & 600\\
4756.683440&19.98896&0.00543 & 600\\
4756.691345&19.97164&0.00657 & 600\\
4756.699261&19.95174&0.00564 & 600\\
4756.707640&19.95588&0.00544 & 600\\
4756.715788&19.95272&0.00518 & 600\\
4756.723530&19.94652&0.00542 & 600\\
4756.731748&19.94944&0.00645 & 600\\
4756.739965&19.94136&0.00637 & 600\\
4756.747869&19.93104&0.00559 & 600\\
4756.755936&19.93294&0.00689 & 600\\
4756.764084&19.92913&0.00640 & 600\\
4756.772069&19.90027&0.00690 & 600\\
4756.780067&19.90065&0.00678 & 600\\
4756.788295&19.88500&0.00731 & 600\\
\\
\hline
\end{tabular}
\end{table}

\begin{table}
\caption{RV data for WASP-15b.}\label{tab:WASP15data}
\begin{tabular}{lllll}
\hline
\hline
bjd (-\,2\,450\,000)&RV (km\,s$^{-1}$)& $\sigma_\mathrm{RV}$ (km\,s$^{-1}$)& $t_\mathrm{exp}$ (s)\\
\hline

\\
CORALIE &\multicolumn{2}{l}{\textit{orbital Doppler shift}}\\
\\
4531.814597&-2.26646&0.01478 & 1801\\
4532.722143&-2.35660&0.01795 & 1801\\
4533.746849&-2.33893&0.01512 & 1801\\
4534.877800&-2.24444&0.01510 & 1801\\
4535.734111&-2.27377&0.01376 & 1801\\
4536.666620&-2.37725&0.01261 & 1801\\
4537.780541&-2.33201&0.00992 & 1801\\
4538.745908&-2.25366&0.01081 & 1801\\
4556.794821&-2.27983&0.01122 & 1801\\
4557.748799&-2.24646&0.01339 & 1801\\
4558.733390&-2.35678&0.01057 & 1801\\
4559.747899&-2.35637&0.01085 & 1801\\
4560.615839&-2.27820&0.01189 & 1801\\
4589.658970&-2.39063&0.01261 & 1801\\
4591.635494&-2.26077&0.01184 & 1801\\
4655.468928&-2.25970&0.01109 & 1801\\
4656.516477&-2.34362&0.01056 & 1801\\
4657.613442&-2.28634&0.01558 & 1801\\
4662.520488&-2.24271&0.00960 & 1801\\
4663.587927&-2.32780&0.01314 & 1801\\
4664.593196&-2.35564&0.01468 & 1801\\
4834.855119&-2.25263&0.01138 & 1801\\
4835.854025&-2.29142&0.01198 & 1801\\
4840.827799&-2.34679&0.01314 & 1801\\
4859.843943&-2.35241&0.01213 & 1801\\
4881.821230&-2.36013&0.01160 & 1801\\
4882.776419&-2.33694&0.01217 & 1801\\
4884.763791&-2.31248&0.01155 & 1801\\
4889.711779&-2.36367&0.01480 & 1801\\
4890.844581&-2.28402&0.01194 & 1801\\
4891.761831&-2.24770&0.01257 & 1801\\
4939.677986&-2.26803&0.01438 & 1801\\
4943.706114&-2.26328&0.01040 & 1801\\
4945.776216&-2.38071&0.01152 & 1801\\
4947.626310&-2.25749&0.01094 & 1801\\
4948.804753&-2.33422&0.01136 & 1801\\
4949.833876&-2.39783&0.01413 & 1801\\
4971.575204&-2.34588&0.02533 & 1801\\
4973.614431&-2.25744&0.01122 & 1801\\
4975.565332&-2.33745&0.01238 & 1801\\
4983.719667&-2.33892&0.01251 & 1801\\
4995.532602&-2.30644&0.01228 & 1801\\
4996.512783&-2.24056&0.01272 & 1801\\
5011.611762&-2.24270&0.02538 & 1801\\
\\
HARPS&\multicolumn{2}{l}{\textit{Rossiter-McLaughlin effect}}\\
\\
4947.563543&-2.23262&0.00254 & 1800\\
4947.812040&-2.22718&0.00256 & 1800\\
4948.542784&-2.27576&0.00475 & 600\\
4948.549428&-2.28213&0.00535 & 500\\
4948.555099&-2.29438&0.00598 & 400\\
4948.560029&-2.28980&0.00601 & 400\\
4948.565111&-2.28430&0.00636 & 400\\
4948.570192&-2.28446&0.00600 & 400\\
4948.575134&-2.28779&0.00585 & 400\\
4948.580168&-2.29174&0.00598 & 400\\
4948.585261&-2.29353&0.00592 & 400\\
4948.590284&-2.30613&0.00571 & 400\\
4948.595319&-2.29820&0.00582 & 400\\
4948.600400&-2.29954&0.00600 & 400\\
4948.605435&-2.29201&0.00570 & 400\\
4948.610423&-2.29414&0.00585 & 400\\
4948.615423&-2.28089&0.00602 & 400\\
4948.620458&-2.28480&0.00607 & 400\\
4948.625539&-2.28205&0.00590 & 400\\
4948.630481&-2.29198&0.00621 & 400\\
4948.635562&-2.28189&0.00646 & 400\\
4948.640609&-2.27422&0.00615 & 400\\
4948.645690&-2.29462&0.00599 & 400\\
4948.650713&-2.27658&0.00615 & 400\\
4948.655701&-2.26646&0.00636 & 400\\
4948.660747&-2.28384&0.00643 & 400\\
4948.665724&-2.26487&0.00670 & 400\\
4948.670852&-2.27511&0.00731 & 400\\
4948.675886&-2.27851&0.00701 & 400\\
4948.680967&-2.26950&0.00638 & 400\\
4948.685921&-2.26493&0.00605 & 400\\
4948.690863&-2.26981&0.00596 & 400\\
4948.695944&-2.28244&0.00585 & 400\\
4948.701083&-2.26951&0.00609 & 400\\
4948.706083&-2.27611&0.00591 & 400\\
4948.711072&-2.25143&0.00605 & 400\\
4948.716141&-2.27273&0.00633 & 400\\
4948.721188&-2.29240&0.00623 & 400\\
4948.726269&-2.31018&0.00619 & 400\\
4948.731245&-2.31258&0.00684 & 400\\
4948.736326&-2.29523&0.00650 & 400\\
4948.741361&-2.31158&0.00604 & 400\\
4948.746350&-2.30475&0.00604 & 400\\
4948.751396&-2.29752&0.00608 & 400\\
4948.757009&-2.30602&0.00560 & 500\\
4948.763641&-2.31262&0.00522 & 600\\
4948.770783&-2.29795&0.00606 & 600\\
4948.778572&-2.30931&0.00609 & 600\\
4948.831697&-2.31076&0.00271 & 1800\\
4949.553111&-2.36641&0.00249 & 1800\\
4949.803656&-2.34899&0.00255 & 1800\\

\\
\hline
\end{tabular}
\end{table}

\begin{table}
\caption{RV data for WASP-17b.}\label{tab:WASP17data}
\begin{tabular}{lllll}
\hline
\hline
bjd (-\,2\,450\,000)&RV (km\,s$^{-1}$)& $\sigma_\mathrm{RV}$ (km\,s$^{-1}$) & $t_\mathrm{exp}$ (s)\\
\hline

\\
CORALIE &\multicolumn{2}{l}{\textit{orbital Doppler shift}}\\
\\
4329.603717&-49.45704&0.04276 & 1801\\
4360.486284&-49.36606&0.04444 & 1801\\
4362.497988&-49.51747&0.04074 & 1801\\
4364.487972&-49.48913&0.04321 & 1801\\
4367.488333&-49.44153&0.03428 & 1801\\
4558.883883&-49.49882&0.03107 & 1801\\
4559.770757&-49.57983&0.03246 & 1801\\
4560.731432&-49.57336&0.02950 & 1801\\
4588.779923&-49.48810&0.02888 & 1801\\
4591.777793&-49.46606&0.03401 & 1801\\
4622.691662&-49.39756&0.03508 & 1801\\
4624.636716&-49.44937&0.03672 & 1801\\
4651.619514&-49.45639&0.03187 & 1801\\
4659.524574&-49.46930&0.04045 & 1801\\
4664.642452&-49.54237&0.03533 & 1801\\
4665.659284&-49.49053&0.03768 & 1801\\
4682.582361&-49.50072&0.03086 & 1801\\
4684.626422&-49.51688&0.03572 & 1801\\
4685.514523&-49.47409&0.03067 & 1801\\
4690.618201&-49.57756&0.03503 & 1801\\
4691.607737&-49.51399&0.04058 & 1801\\
4939.845725&-49.43937&0.03486 & 1801\\
4940.734595&-49.58375&0.03094 & 1801\\
4941.851984&-49.54081&0.02903 & 1801\\
4942.695907&-49.47747&0.02271 & 2701\\
4942.874690&-49.41960&0.02908 & 2701\\
4943.665531&-49.51029&0.02530 & 2701\\
4943.887221&-49.58850&0.02681 & 2701\\
4944.685781&-49.55404&0.02498 & 2701\\
4944.868869&-49.57455&0.02493 & 2701\\
4945.696896&-49.54421&0.02523 & 2701\\
4945.827736&-49.57673&0.02625 & 2701\\
4946.728868&-49.46616&0.02511 & 2701\\
4946.906874&-49.45777&0.02557 & 2701\\
4947.655776&-49.50278&0.02608 & 2701\\
4947.869385&-49.50923&0.02514 & 2701\\
4948.641504&-49.52028&0.02507 & 2701\\
4948.883575&-49.62500&0.02543 & 2701\\
4949.864625&-49.46432&0.02787 & 2701\\
4951.666090&-49.52332&0.02495 & 2701\\
4951.871943&-49.54578&0.02712 & 2701\\
4982.719638&-49.57744&0.04394 & 1801\\
4983.784326&-49.47861&0.02903 & 2701\\
4985.641154&-49.55333&0.02307 & 2701\\
4995.733890&-49.45198&0.02959 & 2701\\
5003.534092&-49.55061&0.06991 & 2701\\
5003.567956&-49.61140&0.10818 & 2701\\
5010.654298&-49.50113&0.04273 & 2701\\
5013.621393&-49.46994&0.02968 & 2701\\
\\
CORALIE & \multicolumn{2}{l}{\textit{Rossiter-McLaughlin effect}}\\
\\
4972.771828&-49.45006&0.02583 & 2701\\
4973.656162&-49.49751&0.03027 & 1801\\
4973.681857&-49.48758&0.03160 & 1801\\
4973.720561&-49.54121&0.03338 & 1801\\
4973.743917&-49.52875&0.03615 & 1801\\
4973.767227&-49.52131&0.03342 & 1801\\
4973.793073&-49.51045&0.03325 & 1801\\
4973.816371&-49.45089&0.03254 & 1801\\
4973.839647&-49.43288&0.03681 & 1801\\
4973.863142&-49.38687&0.03440 & 1801\\
4973.886545&-49.37810&0.03488 & 1801\\
4973.909809&-49.52989&0.05071 & 1801\\
4974.734674&-49.57918&0.02560 & 2701\\
4975.619248&-49.40991&0.03130 & 2701\\
4976.737444&-49.46715&0.02672 & 2701\\
\\
HARPS&\multicolumn{2}{l}{\textit{orbital Doppler shift}}\\
\\
4564.819485&-49.48835&0.01084 & 1200\\
4565.873135&-49.43558&0.00919 & 1200\\
4567.851637&-49.53681&0.01048 & 1200\\
5021.538209&-49.45221&0.00909 & 1800\\
5021.745591&-49.44863&0.00927 & 1800\\
5022.528339&-49.50945&0.00927 & 1800\\
5022.713466&-49.51810&0.00853 & 1800\\
5023.583129&-49.56456&0.02112 & 1800\\
5024.563940&-49.47470&0.01130 & 1800\\
5032.648408&-49.42672&0.00949 & 1800\\
5038.689537&-49.51101&0.01112 & 1800\\
5040.673292&-49.46616&0.01437 & 1800\\
5041.637631&-49.52676&0.00868 & 1800\\
5042.493598&-49.50537&0.00847 & 1800\\
5042.661500&-49.52931&0.00767 & 1800\\
5045.502743&-49.55862&0.01085 & 1800\\
\\
HARPS&\multicolumn{2}{l}{\textit{Rossiter-McLaughlin effect}}\\
\\
5018.475525&-49.47606&0.01098 & 900\\
5018.486520&-49.46933&0.01103 & 900\\
5018.496750&-49.44767&0.01496 & 600\\
5018.504065&-49.48909&0.01522 & 600\\
5018.511379&-49.45537&0.01597 & 600\\
5018.518694&-49.45999&0.01449 & 600\\
5018.526008&-49.46642&0.01478 & 600\\
5018.533392&-49.49716&0.01447 & 600\\
5018.540579&-49.48659&0.01459 & 600\\
5018.547974&-49.48097&0.01570 & 600\\
5018.555358&-49.51021&0.01576 & 600\\
5018.562476&-49.50083&0.01599 & 600\\
5018.569940&-49.56032&0.01673 & 600\\
5018.577174&-49.54420&0.01915 & 600\\
5018.584615&-49.53700&0.01804 & 600\\
5018.591791&-49.55900&0.01609 & 600\\
5018.599105&-49.56062&0.01586 & 600\\
5018.606350&-49.53495&0.01772 & 600\\
5018.613815&-49.46437&0.02020 & 600\\
5018.621129&-49.48921&0.01823 & 600\\
5018.628444&-49.45563&0.01820 & 600\\
5018.635839&-49.45895&0.01796 & 600\\
5018.643084&-49.41901&0.01877 & 600\\
5018.650387&-49.37979&0.02015 & 600\\
5018.657689&-49.41304&0.01889 & 600\\
5018.664807&-49.34699&0.02013 & 600\\
5018.672619&-49.35611&0.02084 & 600\\
5018.679598&-49.39587&0.01887 & 600\\
5018.687120&-49.38167&0.01758 & 600\\
5018.694157&-49.37674&0.01644 & 600\\
5018.701610&-49.37749&0.01821 & 600\\
5018.709526&-49.42154&0.01961 & 600\\
5018.716632&-49.45351&0.02019 & 600\\
5018.723808&-49.46651&0.02567 & 600\\
5018.731713&-49.46214&0.02757 & 600\\
5018.739177&-49.51846&0.02501 & 600\\
5018.746214&-49.47714&0.02148 & 600\\
5018.753459&-49.57730&0.02218 & 600\\
5018.760715&-49.52289&0.02544 & 600\\
5018.768180&-49.45964&0.02967 & 600\\
5018.777034&-49.42103&0.02559 & 900\\
5018.787508&-49.46160&0.03432 & 900\\
5020.504678&-49.51044&0.00829 & 1800\\
5020.724131&-49.47401&0.01131 & 1800\\
\\
\hline
\end{tabular}
\end{table}

\begin{table}
\caption{RV data for WASP-18b.}\label{tab:WASP18data}
\begin{tabular}{lllll}
\hline
\hline
bjd (-\,2\,450\,000)&RV (km\,s$^{-1}$)& $\sigma_\mathrm{RV}$ (km\,s$^{-1}$)& $t_\mathrm{exp}$ (s)\\
\hline

\\
CORALIE&\multicolumn{2}{l}{\textit{orbital Doppler shift}}\\
\\
4359.815630&4.03853&0.01401 & 1801\\
4362.673973&3.71230&0.01276 & 1177\\
4363.733268&2.28379&0.01041 & 1177\\
4655.938244&3.03859&0.00835 & 1801\\
4657.938708&4.40045&0.01057 & 1801\\
4658.892224&4.52777&0.01116 & 1801\\
4660.935178&5.11910&0.00931 & 1801\\
4661.926785&4.85226&0.00919 & 1801\\
4662.911111&4.53095&0.00918 & 1801\\
4760.700356&5.15668&0.00851 & 1801\\
4762.730774&4.29008&0.00887 & 1801\\
4767.543780&3.01935&0.01029 & 1801\\
4767.675234&1.81050&0.00839 & 1801\\
4767.845516&1.82182&0.01167 & 1801\\
4769.805218&2.49459&0.01015 & 1801\\
4770.576633&1.52153&0.01416 & 1801\\
4770.715597&2.15763&0.00954 & 1801\\
4772.648582&2.66395&0.00907 & 1801\\
4772.751819&3.93119&0.00969 & 1801\\
4773.599640&2.78436&0.00930 & 1801\\
4774.606031&3.60566&0.00918 & 1801\\
4775.655139&4.72774&0.00966 & 1801\\
4776.562493&4.44656&0.01098 & 1801\\
4777.543338&4.74172&0.01146 & 1801\\
4778.581020&5.15685&0.00902 & 1801\\
4779.621363&4.78250&0.01012 & 1801\\
4780.551063&4.85945&0.01085 & 1801\\
4781.617770&3.73935&0.00824 & 1801\\
4782.631526&2.78257&0.00871 & 1801\\
4783.635028&2.14893&0.00884 & 1801\\
4825.570049&4.84588&0.00972 & 1801\\
4827.645241&4.71715&0.00915 & 1801\\
4831.640624&2.24706&0.00887 & 1801\\
4836.591194&1.91000&0.00975 & 1801\\
4838.557763&2.78064&0.01005 & 1801\\
4854.571845&2.89455&0.00919 & 1801\\
4857.590403&4.91735&0.01008 & 1801\\
\\
HARPS&\multicolumn{2}{l}{\textit{Rossiter-McLaughlin effect}}\\
\\
4699.683362&3.97402&0.00923 & 600\\
4699.690445&3.90017&0.00802 & 600\\
4699.708200&3.80314&0.01065 & 600\\
4699.716083&3.69215&0.01002 & 600\\
4699.723467&3.57153&0.00931 & 600\\
4699.730400&3.48297&0.00851 & 600\\
4699.738236&3.35447&0.00777 & 600\\
4699.745817&3.21381&0.00615 & 600\\
4699.752994&3.09150&0.00611 & 600\\
4699.760517&2.99338&0.00632 & 600\\
4699.769476&2.89795&0.00584 & 600\\
4699.776640&2.83815&0.00550 & 600\\
4699.783967&2.80582&0.00571 & 600\\
4699.791282&2.73026&0.00655 & 600\\
4699.798863&2.64994&0.00639 & 600\\
4699.806028&2.55609&0.00557 & 600\\
4699.833413&2.28105&0.00479 & 600\\
4699.858483&2.05895&0.00499 & 600\\
4699.917420&1.65925&0.00470 & 600\\
4702.913698&2.02102&0.00544 & 600\\
4704.818564&2.22501&0.00528 & 300\\
4706.792686&3.25812&0.00582 & 1800\\
4709.781421&4.92900&0.00441 & 677\\
\\
\hline
\end{tabular}
\end{table}

\end{document}